\documentclass{article}

\usepackage{arxiv}

\usepackage{lipsum}

\newcommand\blfootnote[1]{%
  \begingroup
  \renewcommand\thefootnote{}\footnote{#1}%
  \addtocounter{footnote}{-1}%
  \endgroup
}

\usepackage{caption}
\usepackage{subcaption}
\usepackage{xcolor}

\usepackage{enumitem} 
\usepackage{bbm} 
\usepackage{xurl} 
\usepackage{tablefootnote}
\definecolor{RCodeColor}{RGB}{248,248,248}
\usepackage{etoolbox}

\usepackage{enumitem} 
\usepackage{bbm} 
\usepackage{xurl} 
\usepackage{tablefootnote}

\usepackage{listings}
\usepackage{xcolor}  

\usepackage{amssymb,amsmath}
\usepackage{ifxetex,ifluatex} 
\ifnum 0\ifxetex 1\fi\ifluatex 1\fi=0 
  \usepackage[T1]{fontenc}
  \usepackage[utf8]{inputenc}
\else 
  \ifxetex
    \usepackage{mathspec}
  \else
  \fi
  \defaultfontfeatures{Ligatures=TeX,Scale=MatchLowercase}
\fi
\IfFileExists{upquote.sty}{\usepackage{upquote}}{}
\IfFileExists{microtype.sty}{%
\usepackage[]{microtype}
\UseMicrotypeSet[protrusion]{basicmath} 
}{}
\PassOptionsToPackage{hyphens}{url} 
\usepackage{hyperref}
\hypersetup{
            pdftitle={Bayesian integrative factor approaches, with an application on nutrition and genomics data},
            pdfborder={0 0 0},
            breaklinks=true}
\usepackage{geometry}
\usepackage{color}
\usepackage{fancyvrb}
\usepackage{framed}
\usepackage{graphicx,grffile}
\makeatletter
\def\maxwidth{\ifdim\Gin@nat@width>\linewidth\linewidth\else\Gin@nat@width\fi}
\def\maxheight{\ifdim\Gin@nat@height>\textheight\textheight\else\Gin@nat@height\fi}
\makeatother
\setkeys{Gin}{width=\maxwidth,height=\maxheight,keepaspectratio}
\IfFileExists{parskip.sty}{%
\usepackage{parskip}
}{
\setlength{\parindent}{0pt}
\setlength{\parskip}{6pt plus 2pt minus 1pt}
}
\ifx\paragraph\undefined\else
\let\oldparagraph\paragraph
\renewcommand{\paragraph}[1]{\oldparagraph{#1}\mbox{}}
\fi
\ifx\subparagraph\undefined\else
\let\oldsubparagraph\subparagraph
\renewcommand{\subparagraph}[1]{\oldsubparagraph{#1}\mbox{}}
\fi

\makeatletter
\def\fps@figure{htbp}
\makeatother
\usepackage{booktabs}
\usepackage{longtable}
\usepackage{array}
\usepackage{multirow}
\usepackage{wrapfig}
\usepackage{float}
\usepackage{colortbl}
\usepackage{pdflscape}
\usepackage{tabu}
\usepackage{threeparttable}
\usepackage{threeparttablex}
\usepackage{ulem}
\usepackage{makecell}

\newcommand\BibTeX{{\rmfamily B\kern-.05em \textsc{i\kern-.025em b}\kern-.08em
T\kern-.1667em\lower.7ex\hbox{E}\kern-.125emX}}

\title{Bayesian integrative factor analysis methods, with application in nutrition and genomics data}

\author{ 
\href{https://orcid.org/0000-0002-8762-7396}{\includegraphics[scale=0.06]{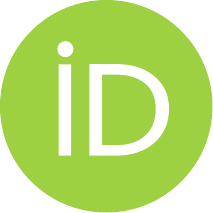}\hspace{1mm}
  Mavis Liang} \\
	Brown University\\
	\texttt{xinwen\_liang@brown.edu} \\  
    \And
	\href{https://orcid.org/0009-0004-3031-6409}{\includegraphics[scale=0.06]{orcid.pdf}\hspace{1mm}Blake Hansen} \\
	Brown University\\
  \texttt{blake\_hansen@brown.edu}\\
  \And
	\href{https://orcid.org/0000-0001-5314-1681}{\includegraphics[scale=0.06]{orcid.pdf}\hspace{1mm}Alejandra Avalos-Pacheco}* \\
	Johannes Kepler University Linz\\
  Harvard Medical School\\
  \texttt{alejandra.avalos@tuwien.ac.at}
    \And
	\href{https://orcid.org/0000-0003-0639-5341}{\includegraphics[scale=0.06]{orcid.pdf}\hspace{1mm}Roberta De Vito}* \\
	 Sapienza University of Rome\\
    Brown University\\
  \texttt{roberta.devito@uniroma1.it}\\
}

\begin{document}
\maketitle













\begin{abstract}
High‐dimensional data are crucial in biomedical research. Integrating such data from multiple studies is a critical process that relies on the choice of advanced statistical models, enhancing statistical power, reproducibility, and scientific insight compared to analyzing each study separately.  
Factor analysis (FA) is a core dimensionality reduction technique that  models observed data through a small set of latent factors. Bayesian extensions of FA have recently emerged as powerful tools for multi-study integration, enabling researchers to disentangle shared biological signals from study-specific variability. In this tutorial, we provide a practical and comparative guide to five advanced Bayesian integrative factor models:  Perturbed Factor Analysis (PFA), Bayesian Factor Regression with non-local spike‐and‐slab priors (MOM-SS), Subspace Factor Analysis (SUFA), Bayesian Multi-study Factor Analysis (BMSFA), and Bayesian Combinatorial Multi-study Factor Analysis (Tetris). To contextualize these methods, we also include two benchmark approaches: standard FA applied to pooled data (Stack FA) and FA applied separately to each study (Ind FA). We evaluate all methods through extensive simulations, assessing computational efficiency, accuracy in estimation of loadings and numbers of factors. 
To bridge theory and practice, we present a full analytical workflow---with detailed R code---demonstrating how to apply these models to real-world datasets in nutrition and genomics.
This tutorial is designed to guide applied researchers through the landscape of Bayesian integrative factor analysis, offering insights and tools for extracting interpretable, robust patterns from complex multi-source data. All code and resources are available at: \href{https://github.com/Mavis-Liang/Bayesian_integrative_FA_tutorial}{https://github.com/Mavis-Liang/Bayesian\_integrative\_FA\_tutorial}.
\end{abstract}
 



\blfootnote{* These authors contributed equally to this work}

\section{Introduction}

The rapid evolution of high-throughput biological technologies has transformed biomedical research by producing large-scale, complex, and heterogeneous data sets in different studies, platforms, and populations. 
Developing scalable and interpretable methods for their integrative analysis is critical to capture deep biological insights from such data, but their complexity demands advanced statistical frameworks that can go beyond standard approaches \cite{mardis2013next, hu2021next}.
 Integration of such different studies is essential to improve statistical power, estimate reproducibility, and facilitate the discovery of robust biological signals by accounting for shared structures between datasets while modeling study-specific variation \cite{leek2010tackling, garrett2008cross}. However, integrating data from diverse origins is far from trivial: batch effects, platform-specific biases, and population heterogeneity introduce challenges that require sophisticated statistical modeling \cite{garrett2008cross}.

Dimensionality reduction techniques—such as Principal Component Analysis (PCA) and Factor Analysis (FA)—are foundational tools in high-dimensional data analysis. These methods reduce complexity, facilitate visualization, and identify important latent structures underlying observed variables \cite{cattell2012scientific}. In particular, Bayesian formulations of FA have proven powerful in high-dimensional contexts: by leveraging priors that induce sparsity or shrinkage, such as spike-and-slab priors \cite{george1993variable, carvalho2008high, rovckova2016fast}, cumulative shrinkage priors \cite{legramanti2020bayesian}, or generalized sparse priors \cite{schiavon2022generalized}: Bayesian FA enables more interpretable latent representations. 
These models also offer principled uncertainty quantification and automatic determination of the number of latent factors, making them attractive for complex biological applications \cite{lopes2004bayesian}.

Despite their utility, standard FA and PCA are not designed to handle the complexity of multi-study data, where both shared and study-specific sources of variation must be disentangled. Simple approaches like pooling data across studies (``stacking'') or analyzing each study separately (``independent analysis'') often obscure critical signals or amplify spurious artifacts, leading to misleading conclusions.
In many modern applications—such as integrating gene expression data from different microarray platforms \cite{wang2011unifying}, harmonizing nutritional data from case-control studies across populations \cite{edefonti2012nutrient}, or modeling imaging data collected from multiple subjects and time points \cite{shinohara2014statistical}—there is a clear need for statistical methods that can explicitly separate shared biological structure from study-specific technical or biological variation.

To meet this need, multi-study extensions of Bayesian FA have emerged. Multi-Study Factor Analysis (MSFA) \cite{de2019multi} and its Bayesian counterpart (BMSFA) \cite{de2021bayesian} jointly model common and study-specific factors, using shrinkage priors to stabilize estimation in high dimensions.  Perturbed Factor Analysis (PFA) \cite{Roy2021PfaA} allows for structured study-specific deviations via perturbation matrices. Bayesian Latent Factor Regression (e.g., MOM-SS, Laplace-SS) \cite{Avalos2022HLDI} combines spike-and-slab priors with batch adjustment mechanisms to isolate true biological signals. Subspace Factor Analysis (SUFA) \cite{chandra2024inferring} improves identifiability of shared and unique latent spaces, while Bayesian Combinatorial Multi-Study FA (Tetris) \cite{grabski2020combinatorialmsfa} generalizes BMSFA by allowing flexible patterns of factor sharing across studies.

Although these methods offer remarkable flexibility and modeling power, they differ in terms of their assumptions, estimation strategies, and suitability for different types of high-dimensional data. For applied researchers, it is often unclear which model to choose, how to interpret its output, or how to implement it correctly. Existing tutorials in the literature are often domain-specific—for instance, focusing on genomic data and machine learning techniques \cite{mardoc2024genomic}, general data integration principles \cite{doan2012principles}, or nutrition-focused multi-population studies \cite{edefonti2020reproducibility}. However, none of these provide a comprehensive, statistically grounded tutorial focused on Bayesian integrative factor models for multi-study high-dimensional data.
To fill this gap, we develop this tutorial to systematically evaluate and compare the most widely used Bayesian multi-study factor models on a unified overview of these approaches. We provide theoretical and practical guidance on model assumptions, identifiability, computational considerations, and post-processing. In doing so, we help researchers choose the most appropriate method based on their data characteristics and scientific objectives. Our tutorial is structured to support both conceptual understanding and direct application.

We focus on five of the most prominent and cited Bayesian integrative factor models—PFA, MOM-SS, SUFA, BMSFA, and Tetris—alongside two benchmark approaches: standard FA on pooled data (Stack FA) and standard FA applied separately to each study (Ind FA). We compare these methods using extensive simulation studies that vary the number of studies, dimensionality, signal-to-noise ratio, and factor-sharing structure. In addition, we demonstrate real-world applications using genomic and nutritional datasets, with fully reproducible R code to guide implementation.

The remainder of the paper is organized as follows. In Section~\ref{sec:2}, we present the modeling framework for multi-study factor analysis, detail each Bayesian method, and discuss identifiability and inference. Section~\ref{sec:3} presents simulation studies comparing model performance across different settings. Section~\ref{sec:4} illustrates the application of each method to real-world datasets, with full analytical workflows and interpretation strategies. Finally, Section~\ref{sec:5} concludes with practical recommendations and future directions.

\section{Data Integration Methods}\label{sec:2}

\subsection{The Multi-study Data}\label{sec:multistudy data}

\begin{figure}[!ht]
    \centering
    \includegraphics[width=\linewidth]{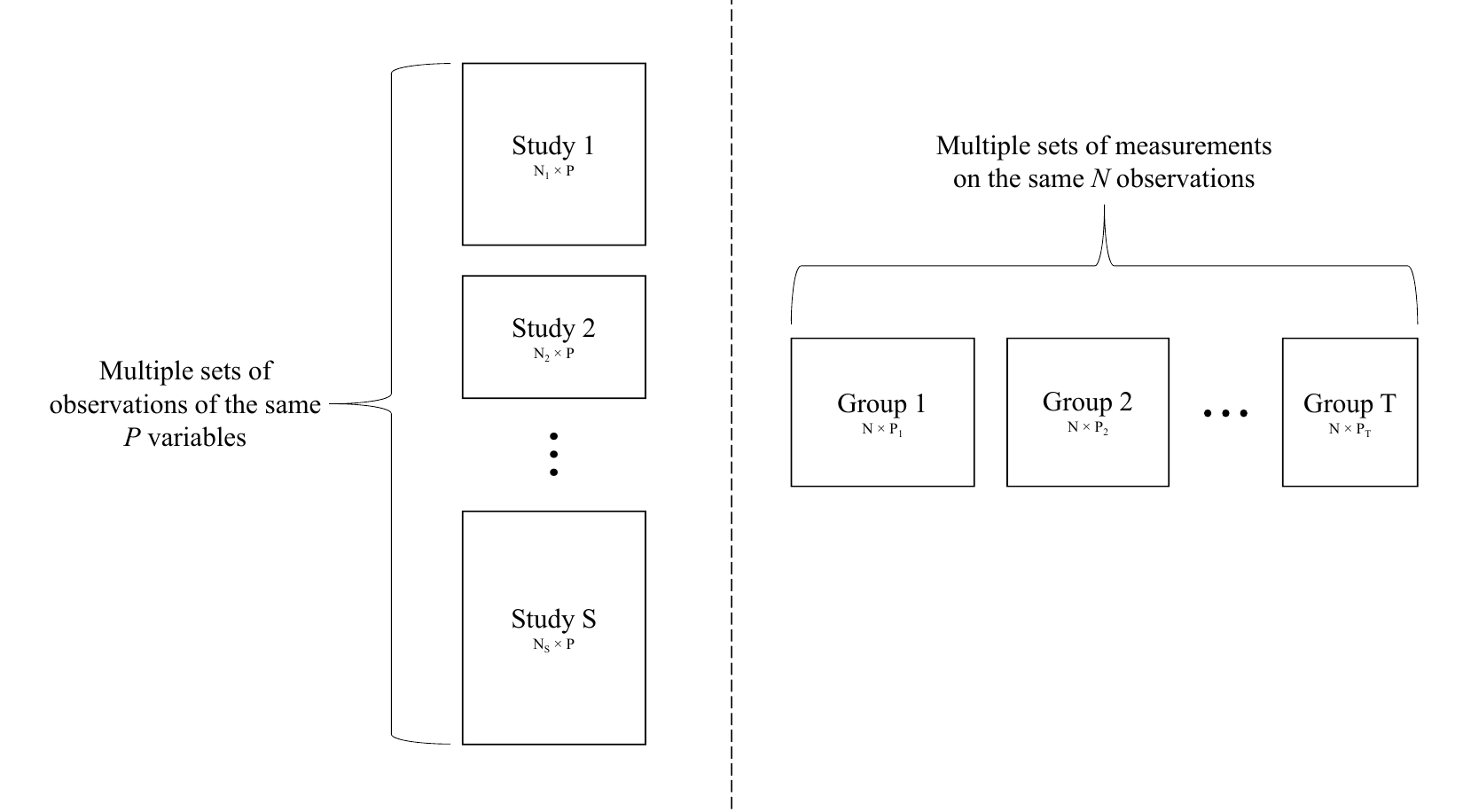}
    \caption{\it Data Integration Scenarios: multi-study data which observes the same set of variables across multiple groups of subjects (left), multi-platform data which observes multiple sets of variables on the same subjects (right).}
    \label{fig:data-integration}
\end{figure}

In this work, we focus on data integration techniques that provide insights of the data that single source analyses can miss. 
Specifically, there are two main model-based data integration problems, as illustrated in Figure \ref{fig:data-integration}: 1.) multi-study data, where the same set of variables is measured across multiple independent studies or groups and 2.)  multi-platform (or multi-modality) data,  where different types of variables are collected on the same set of individuals.

When analyzing multi-study data, the goal of analysis is two-fold: to identify common signal present within all of the studies, as well as to identify study-specific signal unique to each study, or a combination of both. For example, in genomic studies, the aim is to identify robust reproducible biological pathways, that may be masked by batch effects in single-study analyses.  Integrating data across multiple cohorts can enhance statistical power and reveal robust shared components. Conversely,  in some contexts such as nutritional epidemiology, the study-specific effects—like dietary patterns specific to ethnic groups—are of primary interest.

The analysis of multi-platform data aims to quantify different signals between distinct data sources and explore the intricacies of interconnections between those multiple layers. For instance, in oncology studies, multi-modality data including electronic health records, molecular data and medical images are often collected for the individuals\cite{steyaert2023multimodal}. While biomarker discoveries are mainly based on single modality molecular data, integrating different data modalities can improve disease classification and inform precision medicine strategies \cite{steyaert2023multimodal,cheerla2019deep}. Multi-platform data present a different challenge, as we can no longer assume that the data are i.i.d. because the same set of subjects are observed under each platform. Consequently, statistical methods tailored for multi-platform integration—such as Multi-Omics Factor Analysis\cite{argelaguet2018multi} or DIABLO \cite{singh2019diablo}—have emerged to address these challenges.

In this tutorial, we focus on methods designed for the multi-study setting, while recognizing that future extensions are needed to unify multi-study and multi-platform approaches under a common framework.

For simplicity, we use the term "multi-study" to refer to datasets where the same set of $P$ variables are measured across $S$ independent groups, studies or platforms. 
However, we emphasize that multi-study can also be samples from different tissues or locations, replicated data collected in separate days or batches, or even experiments under different treatment conditions, which are subject to specific contexts. 

To begin with, we introduce the general notation for the multi-study setting used throughout this tutorial:

\begin{itemize}
    \item $S$: number of studies.
    \item $N_s$: number of observations in the $s$th study, where $s=1, \cdots, S$. 
    \item $P$: number of variables observed in each of the $S$ studies.
    \item $y_{ips}$: value for the $p$th variable of $i$th observation in $s$th study, for $i=1, \cdots, N_s$, $p=1, \cdots, P$, and $s=1,\cdots, S$; with vector notation $\mathbf{y}_{is} = (y_{i1}, \cdots, y_{iP})^\top_s$. 
    \item $\mathbf{Y}_s = (\mathbf{y}_{1s}, \dots, \mathbf{y}_{N_s s})^\top$: the $N_s \times P$ data matrix for study $s$. 
 \end{itemize}   
   
Figure~\ref{fig:data-integration} provides a visual comparison of the multi-study and multi-platform scenarios. In the remainder of this paper, we focus on integrative methods that assume each study provides measurements on the same variables, but across different samples. These settings are prevalent in genomics, nutrition, and other biomedical applications, and require careful statistical modeling to capture both common and study-specific latent structures.

\subsection{Simple Integrative Factor Approach}

Factor analysis (FA) is a widely used technique for modeling high‐dimensional data $\mathbf{y}_i$ through a smaller number of latent variables.
We begin by introducing the standard FA model in the context of a single study, and then describe two naive strategies for applying this model to multi-study data: Stack FA and Ind FA.

The classical factor model assumes:
\(\mathbf{y}_i\in\mathbb{R}^P\) are modeled as:
\begin{equation}\label{eq:fa}
     \begin{array}{c}
       \underbrace{\mathbf{y}_{i}}_{(P \times 1)} = \underbrace{\Phi}_{(P \times K)} \underbrace{\mathbf{f}_{i}}_{(K \times 1)} + \underbrace{\boldsymbol{\epsilon}_{i}}_{(P \times 1)}, 
    \end{array}    
\end{equation}

where \(\Phi\) is the $P \times K$ factor loadings matrix, \(\mathbf{f}_i\) are a $K$-dimensional independent \textit{latent factor} vector, and \(\boldsymbol{\epsilon}_i\) is the \textit{idiosyncratic residual} term. Typically, the model assumes:
\begin{equation}\label{eq:assum1}
    \begin{array}{cc}
          \mathbf{f}_i\sim \text{MVN}(0,I_K)\,,  \quad\boldsymbol{\epsilon}_i \sim \text{MVN}(0, \Psi)\,,
    \end{array}
\end{equation}
where $\Psi=\text{diag}(\psi_1,\cdots, \psi_P)$. Under independence between $\mathbf{f}_i$ and $\boldsymbol{\epsilon}_i$ and properties of the normal distribution, the marginal distribution of $\mathbf{y}_i$ is: 
\begin{equation}
    \mathbf{y}_i\sim\text{MVN}(0, \Phi\Phi^\top+\Psi).
\end{equation}
Thus, the covariance structure is decomposed into a low-rank loading component $\Phi \Phi^\top$, capturing variability explained by the latent factors, and a diagonal matrix $\Psi$, capturing residual variances.

This model can also be expressed element-wise as: 
\begin{equation} y_{ip}=\sum_{k=1}^{K}\phi_{pk} f_{ik}+\boldsymbol{\epsilon}_{ip}\,,
\end{equation}

where the loading $\phi_{pk}$ reflects the influence of factor $k$ on variable $p$. Since the number of factors are less than the number of observed variables, $K \ll P$, FA offers a compact and interpretable representation of high-dimensional data.

Bayesian factor models focus on priors for factor loadings $\Phi$, and residual covaraince matrix $\Psi$. Flexible prior for the loadings include heavy-tailed priors \cite{ghosh2009default}, spike-and-slab priors \cite{george1993variable, carvalho2008high, rovckova2016fast}, and global-local shrinkage priors such as the Multiplicative Gamma Process Shrinkage (MGPS) prior \cite{bhattacharya2011sparse, legramanti2020bayesian}. These priors help control overfitting and enable automatic selection of the number of active factors.  

Posterior inference proceed via Gibbs sampling, Expectation-Maximization, or variational methods. Selecting the number of latent factors $K$ remains a challenging task, with methods ranging from information criteria \cite{bai2002determining} to fully Bayesian learning approaches \cite{lee2002bayesian, lopes2004bayesian, fruhwirth2024sparse}; see \cite{owen2016bi} for a comprehensive review.

\subsubsection{Stack FA}
\paragraph{Model}
The first model for data integration relies on  stacking all the individulas from the different studies into one combined matrix and and fit a single FA model (Stack FA), effectively ignoring study labels. 

For each subject $i=1, \cdots,N_s$ in study $s=1, \cdots, S$, the model is:
\begin{equation}
    \label{eq:stack_factor}
    \begin{array}{c}
       \underbrace{\mathbf{y}_{is}}_{(P \times 1)} = \underbrace{\Phi}_{(P \times K)} \underbrace{\mathbf{f}_{is}}_{(K \times 1)} + \underbrace{\mathbf{\epsilon}_{is}}_{(P \times 1)}, \\[25pt]
    \end{array}    
\end{equation}
where $\mathbf{f}_{is}\sim \text{MVN}(0,I_K)$ are $K$-dimensional latent factors, and $\mathbf{\epsilon}_{is}\sim \text{MVN}(0,\Psi)$ is the idiosyncratic error with $\Psi=\text{diag}(\psi_1,\cdots, \psi_P)$. 
 Because the same $\Phi$ and $\Psi$ are shared across all studies, this model implies:  
\begin{equation}
    \Sigma_1=\Sigma_2=\cdots=\Sigma_S=\Phi\Phi^\top+\Psi.
\end{equation}
Thus, Stack FA assumes that there is no systematic difference between studies. This homogeneous covariance assumption may be appropriate when the generative processes are similar across studies. However, it can mask meaningful heterogeneity if study-specific structure exists.


\paragraph{Priors}
In this tutorial, we use  the MGPS prior on $\Phi$ as proposed by Bhattacharya and Dunson \cite{bhattacharya2011sparse}: 
\begin{equation}
    \label{eq:mgps}
    \begin{array}{c}
        \phi_{pk} \sim \mathcal{N}(0, \omega_{pk}^{-1} \theta_k^{-1})\,, \\
        \omega_{pk} \sim \text{Gam}(\kappa/2, \kappa/2)\,, \\
        \theta_k  = \prod_{l=1}^{k} \delta_l\,, 
        \delta_1 \sim \text{Gam}(a_1, 1)\,, 
        \delta_l \sim \text{Gam}(a_2, 1) \text{ for } l > 1\,,
    \end{array}
\end{equation}
where $\omega_{pk} \sim \text{Gam}(\kappa/2, \kappa/2)$ is called local shrinkage,  and $\theta_k = \prod_{l=1}^{k} \delta_l$ is called global shrinkage , imposing increasingly strong shrinkage on higher‐indexed factors. This hierarchical shrinkage structure penalizes higher-indexed factors more strongly. 

Default hyperparameters are $\kappa = 3$, $a_1 = 2.1$, and $a_2 = 3.1$. For the residual variances, we assume $\psi_p \sim \text{Inv-Gamma}(1, 0.3)$.  This choice are based on the guidelines by Bhattacharya and Dunson\cite{bhattacharya2011sparse} and Durante\cite{durante}.


\paragraph{Estimation}
The parameters in the Stack FA are typically estimated using Gibbs sampling.

 To choose $K$, the standard procedure is the application of the eigenvalue decomposition (EVD) to the covariance matrices and retain the number of factors explaining a threshold level of explained variance (e.g., $\geq$ 5\%).  This is a \textit{post-hoc} strategy recommended by BMSFA. We then run the model again with the estimated numbers of factors.

 Alternatively, Bhattacharya and Dunson proposed a dynamic truncation strategy, in which $K$ is initialized to $3 \log(P)$ and adaptively pruned or expanded during sampling. This approach is implemented in MATLAB.

\subsubsection{Ind FA}

\paragraph{Model}
An alternative approach is to fit separate FA models for each study (Ind FA). Specifically, for study $s$ we model
\begin{equation}
    \label{eq:independent_factor}
    \begin{array}{c}
        \underbrace{\mathbf{y}_{is} }_{(P \times 1)} = \underbrace{\Lambda_s}_{(P \times J_s)} \underbrace{\mathbf{l}_{is}}_{(J_s \times 1)} + \underbrace{\mathbf{\epsilon}_{is}}_{(P \times 1)}, \\[25pt]
    \end{array}
\end{equation}
where $\Lambda_s$ is the $P \times J_s$ loadings matrix specific to study $s$,  $\mathbf{l}_{is} \sim \text{MVN}(0, I_{J_s})$, and $\boldsymbol{\epsilon}_{is} \sim \text{MVN}(0, \Psi_s)$ with $\Psi_s = \text{diag}(\psi_{1s}, \dots, \psi_{Ps})$. The marginal covariance of $\mathbf{y}_{is}$ is then: \begin{equation} \Sigma_s = \Lambda_s \Lambda_s^\top + \Psi_s, \end{equation} which we refer to as the study-specific covariance.

Under Ind FA, any signals discovered in one study are not necessarily reproducible in the others, as Ind FA does not pool information across studies.

\paragraph{Priors}
We again adopt the MGPS prior, now applied independently to each $\Lambda_s$: 
\begin{equation}
    \lambda_{pk}^{(s)}\sim N(0,\omega_{pk}^{(s),-1}\theta_k^{(s),-1}),
\end{equation}
with analogous priors on $\omega_{pk}^{(s)}$ and $\theta_k^{(s)}$, i.e., $\omega_{pk}^{(s),-1}\sim\text{Gamma}(\kappa/2,\kappa/2)$, and $\theta_k^{(s),-1}=\prod_{l=1}^k\delta_l^{(s)}$.

\paragraph{Estimation}
Each study is fit separately via Gibbs sampling.

The number of factors $J_s$ can be selected using the same strategies as in Stack FA---either through eigenvalue decomposition or adaptive truncation schemes.

\subsection{Advanced Integrative Factor Approach}

While Stack FA and Ind FA provide simple strategies for multi-study analysis, they fail to exploit shared structure across studies or to model study-specific variations in a unified framework. To overcome these limitations, several advanced Bayesian integrative factor models have been developed. These models allow for joint modeling of common and study-specific variation in a statistically principled way.

\subsubsection{PFA}

\paragraph{Model}
Perturbed Factor Analysis (PFA) \cite{Roy2021PfaA} builds on Stack FA model (Equation \eqref{eq:fa}) by introducing study-specific perturbation matrices $Q_s$ that account for deviations from a shared loading structure. The model is:
\begin{align}
  \underbrace{Q_s}_{(P\times P)}\,\underbrace{\mathbf{y}_{is}}_{(P\times 1)}
  &= \underbrace{\Phi}_{(P\times K)}\,\underbrace{\mathbf{f}_{is}}_{(K\times 1)}
    \;+\;\underbrace{\boldsymbol{\epsilon}_{is}}_{(P\times 1)}\,,
  \label{eq:pfa_mod}
\end{align}
 where $Q_s \in \mathbb{R}^{P \times P}$ is a transformation matrix for study $s$, with $Q_1 = I_P$ fixed for a reference study. For $s > 1$, $Q_s \sim \mathrm{MN}(I_P, \alpha_Q I_P, \alpha_Q I_P)$ follows a matrix normal distribution centered at the identity matrix with row covariance and column covariance being $\alpha_Q$ times the identity matrix. Thus $\alpha_Q$ is a scale parameter that sets how much $Q_s$ can deviate from the identity, controlling the amount of perturbation away from the reference study.

 Factor scores are modeled as $\mathbf{f}_{is} \sim \text{MVN}(0, V)$, where $V = \text{diag}(\nu_1, \dots, \nu_K)$. Unlike simpler FA models that might fix $V=I_K$, here the heteroscedastic variances $\nu_k$ help address rotation identifiability issue. We can equivalently view $\Phi V^{1/2}$ as the common loadings matrix shared by all studies. The residual term $\boldsymbol{\epsilon}_{is} \sim \text{MVN}(0, \Psi)$, with $\Psi = \text{diag}(\psi_1, \dots, \psi_P)$ shared across studies.


In PFA, the marginal covariance of \(\mathbf{y}_{is}\) for each study is: \(\Sigma_s=Q_s^{-1}(\Phi V \Phi^\top + \Psi){(Q_s^{-1})}^\top\), with the common covariance defined as \(\Sigma_{\Phi}= \Phi V \Phi^\top +\Psi\). Note that for $s=1$, $\Sigma_1=\Sigma_\Phi$ since \(Q_1=I_P\). Study-specific covariances are then defined as $\Sigma_{\Lambda_s} = \Sigma_s - \Sigma_\Phi$.

\paragraph{Priors}
PFA uses the same MGPS prior for $\Phi$ as Stack FA and Ind FA. 
The variance components $\nu_k$ have inverse-gamma priors: $\nu_k \sim \text{IG}(a_v, b_v)$, with the default values $a_\nu=10$ and $b_\nu=0.1$. The residual variances is assumed to have a weakly informative inverse-gamma prior $\psi_p \sim \text{IG}(0.1, 0.1)$ are weakly informative. The perturbation scale $\alpha_Q$ also follows an inverse-gamma prior: $\alpha_Q \sim \text{IG}(0.1, 0.1)$.

\paragraph{Estimation}
All parameters are estimated via MCMC with a Gibbs sampling algorithm.  The parameter $\alpha_Q$ for the perturbation matrix can either be sampled from its inverse-gamma posterior, or specified by the user. 
Due to the presence of $Q_s$, inference in PFA is more computationally intensive than in Stack FA or Ind FA. However, it offers the critical advantage of aligning each study to a common latent structure, thus preserving global patterns while accounting for structured study-specific variations.

To determine the number of factors, PFA employs the adaptive truncation procedure introduced by Bhattacharya and Dunson\cite{bhattacharya2011sparse}. At each iteration, the algorithm imputes the mean absolute value of the loadings in each column of $\Phi$. If all the loading for a given factor falls below a predefined cutoff value (e.g. $10^{-3}$), that column is removed from the model.   As a result,  the numbers of factors decreases monotonically throughout the sampling process, allowing the model complexity to be gradually pruned.
 
When the cutoff is appropriately chosen and sufficient burn-in is used, this adaptive truncation typically stabilizes early in sampling. However convergence to a single value is not guaranteed. In practice, the number of factors may still fluctuate slightly in the latter part of the chain, necessitating post-processing.

\subsubsection{MOM-SS}

\paragraph{Model}
The MOM-SS model extends the Stack FA model by incorporating study-specific intercepts and noise to correct for both additive and multiplicative batch effects, as well as a regression adjustment. The model is formulated as:
\begin{equation}
    \label{eq:mom-ss}
\begin{array}{c}
    \underbrace{\mathbf{y}_{is}}_{(P \times 1)} = \underbrace{\boldsymbol{\alpha}_s}_{(P \times 1)} + \underbrace{\boldsymbol{\beta}}_{(P \times Q)} \underbrace{\mathbf{x}_{is}}_{(Q \times 1)} + \underbrace{\Phi}_{(P \times K)} \underbrace{\mathbf{f}_{is}}_{(K \times 1)} + \underbrace{\boldsymbol{\epsilon}_{is}}_{(P \times 1)} \,,\\[25pt]
\end{array}
\end{equation}
where $\boldsymbol{\alpha}_s \in \mathbb{R}^P$ is a study-specific intercept vector, $\mathbf{x}_{is} \in \mathbb{R}^Q$ is a vector of covariates, $\boldsymbol{\beta} \in \mathbb{R}^{P \times Q}$ is the matrix of regression coefficients, $\Phi \in \mathbb{R}^{P \times K}$ is the common loading matrix, and $\mathbf{f}_{is} \sim \text{MVN}(0, I_K)$ are the latent factors. The error term $\boldsymbol{\epsilon}_{is} \sim \text{MVN}(0, \Psi_s)$ accounts for study-specific noise with $\Psi_s = \text{diag}(\psi_{1s}, \ldots, \psi_{Ps})$.

A matrix form of the model \eqref{eq:mom-ss} stacks all observations across studies:
\begin{equation}
    \label{eq:mom-ss_matrix}
\begin{array}{c}
    \underbrace{\mathbf{Y}}_{(N\times P)} = \underbrace{\mathbf{M}}_{(N \times S)}\underbrace{\mathbf{\alpha}^\top}_{(S \times P)} + \underbrace{\mathbf{X}}_{(N \times Q)} \underbrace{\boldsymbol{\beta}^\top}_{(Q \times P)} + \underbrace{\mathbf{F}}_{(N \times K)} \underbrace{\Phi^\top}_{(K \times P)}  + \underbrace{\mathbf{E}}_{(N \times P)} ,\\[25pt]
    \text{with } \mathbf{M}= \begin{bmatrix}
    \mathbf{1}_{N_1} & \mathbf{0} & \cdots & \mathbf{0} \\
    \mathbf{0} & \mathbf{1}_{N_2} & \cdots & \mathbf{0} \\
    \vdots & \vdots & \ddots & \vdots \\
    \mathbf{0} & \mathbf{0} & \cdots & \mathbf{1}_{N_S}
    \end{bmatrix}, \text{ and } \mathbf{\mathbf{\alpha}} = \begin{bmatrix}
    \alpha_{11} & \alpha_{12} & \cdots & \alpha_{1S} \\
    \alpha_{21} & \alpha_{22} & \cdots & \alpha_{2S} \\
    \vdots & \vdots & \ddots & \vdots \\
    \alpha_{P1} & \alpha_{P2} & \cdots & \alpha_{PS}
    \end{bmatrix},
\end{array}
\end{equation}

where \(\mathbf{M}\) is a block-diagonal matrix encoding group/study membership, \(\mathbf{\alpha}\) contains intercepts for each study and variable, \(\mathbf{X}\) is the matrix of observed covariates, \(\mathbf{F}\) contains latent factors, and \(\mathbf{E}\) contains residuals. The marginal covariance in study $s$ is $\Sigma_s = ( \Phi \Phi^\top + \Psi_s) $.

\paragraph{Priors}
MOM-SS adopts  a moment-based non-local spike-and-slab  (NLP) by Johnson et al. \cite{johnson2010use, johnson2012bayesian} on the entries of $\Phi$ to induce sparsity in the loadings. Each loading $\phi_{pk}$ is modeled by
\begin{equation}
    \label{eq:spike_slab}
    \begin{array}{c}
        \phi_{pk} \sim (1-\gamma_{pk}) f_\text{spike} + \gamma_{pk}f_\text{slab}, \\
        \gamma_{pk} \sim \text{Bernoulli}(\zeta_k),
    \end{array}
\end{equation}
where the \textbf{spike} is a normal density with a small variance $f_\text{spike}=\mathcal{N}(0, \tau_{0})$, and the \textbf{slab} is a moment-penalized normal density, which takes the form $f_\text{slab}(\phi_{pk})=(\phi_{pk}^2 / \tau_1) \mathcal{N}(0, \tau_1)$. Hence, the prior assigns zero density at $\phi_{pk}$ whenever $\gamma_{pk}=1$, so that $\phi_{pk}$ is forced away from zero (the non-local property). The default values for \(\tau_0\) and \(\tau_1\) are equal to 0.026 and 0.28 to distinguish practically correlated factors.
The inclusion probability $\zeta_k \sim \text{Beta}(a_\zeta/k, b_\zeta)$ controls the proportion of active loadings, with defaults $a_\zeta = b_\zeta = 1$.  

Priors for the intercepts $\boldsymbol{\alpha}_s$ and regression coefficients $\boldsymbol{\beta}$ follow a standard normal distribution: \(\mathcal{N}(0,\sigma^2_\text{reg})\) with default  \(\sigma^2_\text{reg}=1\).  Each residual variance $\psi_{ps}$ follows an inverse-gamma prior: $\psi_{ps} \sim \text{IG}(0.5, 0.5)$.

\paragraph{Estimations}
Parameter estimation is performed using the Expectation-Maximization (EM) algorithm, which consists of two iterative steps.
The E-step computes the conditional expectations of the latent factors and the M-step maximizes the complete-data log-posterior, updating $\boldsymbol{\alpha}_s$, $\boldsymbol{\beta}$, $\Phi$, and $\Psi_s$.
 A coordinate descent step is embedded in the M-step to accommodate the non-local slab structure.

To initialize parameters, two approaches based on least-square estimation and eigenvalue decomposition are provided—one using eigenvalue decomposition with varimax rotation and another without. Ten-fold cross-validation is used to select the best initialization.

After estimation, the effective number of factors is determined by discarding columns of $\Phi$ with low posterior inclusion probabilities (i.e., $\gamma_{pk} \approx 0$). Factors are reordered \textit{post hoc} based on the number of non-zero loadings to enhance interpretability.

\subsubsection{SUFA}

\paragraph{Model}
SUbspace Factor Analysis (SUFA)\cite{chandra2024inferring} explicitly models both shared and study-specific variation by augmenting a shared factor structure with study-specific subspaces.  The model is defined as:
\begin{equation}
    \label{eq:sufa}
\begin{array}{c}
    \underbrace{\mathbf{y}_{is}}_{(P \times 1)} = \underbrace{\Phi}_{(P \times K)} \underbrace{\mathbf{f}_{is}}_{(K \times 1)} + \underbrace{\Phi}_{(P \times K)}\underbrace{A_{s}}_{(K \times J_s)} \underbrace{\mathbf{l}_{is}}_{(J_s \times 1)}  + \underbrace{\boldsymbol{\epsilon}_{is}}_{(P \times 1)} \,,\\[25pt]
\end{array}
\end{equation} 
where $\Phi \in \mathbb{R}^{P \times K}$ is a shared loading matrix, $\mathbf{f}{is} \sim \text{MVN}(0, I_K)$ are the corresponding common latent factors, $A_s \in \mathbb{R}^{K \times J_s}$ is a transformation matrix for study-specific components, and $\mathbf{l}_{is} \sim \text{MVN}(0, I_{J_s})$ are study-specific latent factors. The residuals $\boldsymbol{\epsilon}_{is} \sim \text{MVN}(0, \Psi)$ are shared across studies, with $\Psi = \text{diag}(\psi_1, \ldots, \psi_P)$. SUFA decomposes the total variability into (i) a shared low‐dimensional subspace common to all studies and (ii) study‐specific subspaces that explain extra variation unique to each study. The first term $\Phi\mathbf{f}_{is}$ is the common factor effect shared across studies, while $\Phi A_s\mathbf{l}_{is}$ introduces additional latent dimensions unique to the study $s$. In this way, SUFA ensures all study-specific loadings $\Lambda_s = \Phi A_s$ lie in the column space of $\Phi$.   Identifiability is guaranteed by enforcing $\sum_s J_s \leq K$. The residuals $\boldsymbol{\epsilon}_{is}$ follow $\text{MVN}(0,\Psi)$ with $\Psi=\text{diag}(\psi_{1},\cdots, \psi_{P})$ invariant across studies.

 The marginal covariance of the study $s$ is \(\Sigma_s = \Phi \Phi^\top + \Phi A_s A_s^\top \Phi^\top + \Psi\), where $\Sigma_\Phi=\Phi\Phi^\top$ represents the shared variance, while $\Sigma_{\Lambda_s}=\Lambda_s\Lambda_s^\top = \Phi A_sA_s^\top\Phi^\top$ captures study-specific effects in study $s$.

\paragraph{Priors}
To encourage sparsity in the shared loading matrix $\Phi$, SUFA adopts a Dirichlet-Laplace (DL)\cite{bhattacharya2015dirichlet} prior.
 The DL prior combines computational efficiency with strong theoretical properties, such as achieving near-minimax optimal posterior contraction rates in high-dimensional settings \cite{pati2014posterior}.  Formally, the vectorized loading matrix  $\text{vec}(\Phi)$ follows $\text{DL}(a_{DL})$ distribution, defined as: 
\begin{equation}\label{eq:DL}
    \begin{array}{cc}
         \phi_{pk} \sim \text{Laplace}(\omega_{p,DL}\theta_{DL}),  \\
         \mathbf{\omega}_{DL} \sim \text{Dir}(a_{DL}, \cdots, a_{DL})\,,
         \theta_{DL} \sim \text{Gam}(a_{DL}P, 0.5),
    \end{array}
\end{equation}
where $\phi_{pk}$ denotes the $(p,k)$th element of $\Phi$. 
 Each scale parameter $\omega_{p,DL}$ comes from a Dirichlet distribution, controlling local shrinkage across rows, while $\theta_{DL}$ is a global shrinkage parameter drawn from a Gamma distribution. The hyperparameter $a_{DL}$ regulates overall sparsity, and is typically set to $0.5$ by default.

For the study-specific transformation matrices $A_s$, SUFA uses independent Gaussian priors: $a_{pjs} \sim \mathcal{N}(0, \sigma_A^2)$, where \(\sigma_A^2=1\) by default. These priors do not induce shrinkage, but help prevent information switching between elements.

The residual variances $\Psi$  have log-normal priors, i.e., $\psi_p \sim \log\mathcal{N}(\mu_\Psi, \sigma_\Psi^2)$, which are known to improve numerical stability and mixing in hierarchical models \cite{gelman2006prior}. By default, the hyperparameters  \(\mu_\Psi\) and \(\sigma^2_\Psi\) are chosen such that \(\mathbbm{E}(\psi_p)=1\) and var$(\psi_p)=7$ for \(p=1, \cdots, P\).

\paragraph{Estimation}
SUFA employs a hybrid MCMC approach, combining a Hamiltonian Monte Carlo (HMC) sampler \cite{neal2012mcmc} within a Gibbs sampling framework.  Rather than sampling from the full joint likelihood, SUFA marginalizes over the latent factors and conducts inference using the marginal likelihood. In each iteration, the parameters, i.e., $\Phi$, $\Psi$, and $A_s$, $s=1, \cdots, S$, are updated via the HMC sampler, while the DL hyperparameters are sampled from their conditional distributions using standard Gibbs steps. 
SUFA parallelizes the HMC step across studies to improve computational efficiency. However, this comes at the cost of additional gradient computations, with runtime scaling approximately quadratically in the number of variables $P$.

For model selection, SUFA requires pre-specification of a maximum value for the number of shared factors $K$. To prune the $K$ into a desired number, it uses a singular value decomposition (SVD)-based algorithm known as \textit{Implicitly Restarted Lanczos Bidiagonalization} \cite{baglama2005augmented}, selecting the smallest $K$ that explains at least 95\% of the total variability before running the MCMC algorithm.
To specify the number of study-specific factors $J_s$, SUFA provides a default heuristic that sets $J_s = K/S$, evenly distributing the total latent component among studies. 

\subsubsection{BMSFA}

\paragraph{Model}
Bayesian Multi-Study Factor Analysis (BMSFA) extends the classical factor model by simultaneously modeling shared and study-specific structures across multiple datasets.  Initially proposed in a frequentist framework \cite{de2019multi}, a fully Bayesian implementation was later introduced \cite{de2021bayesian}. The BMSFA model is defined as:
\begin{equation}
    \label{eq:bmsfa}
\begin{array}{c}
    \underbrace{\mathbf{y}_{is}}_{(P \times 1)} = \underbrace{\Phi}_{(P \times K)} \underbrace{\mathbf{f}_{is}}_{(K \times 1)} + \underbrace{\Lambda_s}_{(P \times J_s)} \underbrace{\mathbf{l}_{is}}_{(J_s \times 1)} \underbrace{\boldsymbol{\epsilon}_{is}}_{(P \times 1)} \,,
\end{array}
\end{equation}
where $\Phi \in \mathbb{R}^{P \times K}$ is the common loading matrix shared by all studies, $\mathbf{f}_{is} \sim \text{MVN}(0, I_K)$ are the corresponding shared latent factors, $\Lambda_s \in \mathbb{R}^{P \times J_s}$ is the study-specific loading matrix for study $s$, and $\mathbf{l}_{is} \sim \text{MVN}(0, I_{J_s})$ are the study-specific latent factors. The residual term $\boldsymbol{\epsilon}_{is} \sim \text{MVN}(0, \Psi_s)$, with $\Psi_s = \text{diag}(\psi_{1s}, \dots, \psi_{Ps})$.

Under this formulation, the marginal covariance for study $s$ is: $\Sigma_s =\Phi\Phi^\top +\Lambda_s\Lambda_s^\top +\Psi_s,$ where $\Phi\Phi^\top$ represents the shared covariance matrix, and each $\Lambda_s\Lambda_s^\top$ captures additional variability unique to study $s$.
 This dual structure generalizes both Stack FA and Ind FA, allowing for modeling heterogeneity across studies.

\paragraph{Priors}
BMSFA applies the Multiplicative Gamma Process Shrinkage (MGPS) prior \cite{bhattacharya2011sparse} to both the shared loading matrix $\Phi$ and the study-specific loading matrices $\Lambda_s$, encouraging shrinkage. Each loading $\phi_{pk}$ (or $\lambda_{p j}^{(s)}$) is assigned a normal prior with variance governed by local and column-wise shrinkage parameters, as described in Equation \eqref{eq:mgps}. This setup ensures that higher-indexed columns are increasingly shrunk toward zero.
Residual variances $\psi_{ps}$ are modeled independently across studies, using inverse-gamma priors: $\psi_{ps} \sim \text{IG}(a_\psi, b_\psi)$, with default values typically set to $(1, 0.3)$.

\paragraph{Estimations}
BMSFA is estimated using a Gibbs sampling algorithm, with full conditional updates for all model parameters. To obtain the posterior means or medians of $\Phi$ and $\Lambda_s$, the authors recommend either Orthogonal Procrustes (OP) rotation or spectral decomposition (SD) for aligning MCMC draws and improving interpretability.

To determine  the number of factors, one can run the sampler with large initial values of $K$ and $J_s$, then post-process the posterior mean of $\Sigma_s$ using an eigenvalue decomposition, retaining factors that explain a specified fraction of variance \cite{de2021bayesian}.
To determine the number of factors, BMSFA initializes with large values for $K$ and $J_s$, and then post-processes the posterior samples of the shared covariance matrix, $\Phi\Phi^\top$, and study-specific covariance matrix, $\Lambda_s\Lambda_s^\top$, using eigenvalue decomposition. Factors are retained if their associated eigenvalues explain a sufficiently large portion of the variance (e.g. 5\%), following the strategy described in De Vito et al.\cite{de2021bayesian}.

\subsubsection{Tetris}

\paragraph{Model}
Tetris \cite{grabski2023bayesian} generalizes BMSFA by enabling combinatorial of shared, partially shared, and study-specific latent factors across different sets of studies. This is achieved through the introduction of a binary indicator matrix $\mathcal{T} \in {0,1}^{S \times K^*}$, where each row corresponds to a study and each column indicates whether a given factor is active in that study.
 The model is written as:
\begin{equation}
    \label{eq:tetris}
\begin{array}{c}
    \underbrace{\mathbf{y}_{is}}_{(P \times 1)} = \underbrace{\Phi^*}_{(P \times K^*)}\underbrace{T_s}_{(K^* \times K^*)} \underbrace{\mathbf{f}_{is}}_{(K^* \times 1)} + \underbrace{\boldsymbol{\epsilon}_{is}}_{(P \times 1)} \,,\\[25pt]
\end{array}
\end{equation}
where $\Phi^* \in \mathbb{R}^{P \times K}$ is the loading matrix, $T_s \in \{0,1\}^{K^ \times K}$ is a diagonal matrix selecting factors for study $s$, $\mathbf{f}_{is} \sim \text{MVN}(0, I_{K})$, and $\boldsymbol{\epsilon}_{is} \sim \text{MVN}(0, \Psi_s)$ with $\Psi_s = \text{diag}(\psi_{1s}, \dots, \psi_{Ps})$. 

The binary matrix $\mathcal{T}$ governs the structure of $T_s$ for each study. Each row of $\mathcal{T}$ corresponds to a study, and each column corresponds to a latent factor. A factor is considered: (1)- common  if the corresponding column of $\mathcal{T}$ contains all 1s, (2)- study-specific , if the column contains a 1 in only one row, (3)-  partially shared if only some rows indicating 1s.

To illustrate, consider a case with $S = 3$ studies. Suppose that the model estimates the following binary matrices $T_1$, $T_2$, $T_3$, and $\mathcal{T}$:
\begin{equation*}
    \label{eq:tetris_example}
\begin{array}{c}
    T_1 = \begin{bmatrix}
    1 & 0 & 0 & 0 \\
    0 & 1 & 0 & 0 \\
    0 & 0 & 1 & 0 \\
    0 & 0 & 0 & 0
    \end{bmatrix},
    \quad
    T_2 = \begin{bmatrix}
    1 & 0 & 0 & 0\\
    0 & 1 & 0 & 0\\
    0 & 0 & 0 & 0\\
    0 & 0 & 0 & 1
    \end{bmatrix},
    \quad
    T_3 = \begin{bmatrix}
    1 & 0 & 0 & 0\\
    0 & 0 & 0 & 0 \\
    0 & 0 & 0 & 0\\
    0 & 0 & 0 & 0
    \end{bmatrix},
    \quad
    \mathcal{T} = \begin{bmatrix}
    1 & 1 & 1 & 0 \\
    1 & 1 & 0 & 1\\
    1 & 0 & 0 & 0
    \end{bmatrix},
\end{array}
\end{equation*}
In this example: Factor 1 is common to all studies (column of all 1s), Factor 2 is shared between studies 1 and 2, Factor 3 is specific to study 1, Factor 4 is specific to study 2.

The common loadings and study-specific loadings can be extracted from $\Phi^*$ by manipulating the $T_s$ matrices.
We can define $P\in \{0,1\}^{K^* \times K^*}$ a diagonal matrix with 1 in the $k$th row if column $k$ in $\mathcal{T}$ has all 1s. Then the common loadings matrices are $\Phi = \Phi^*P$. For each study $s$, let $R_s=T_s-P$, then $\Lambda_s=\Phi^*R_s$ are the loadings corresponding to the partially shared or study-specific factors used by that study. The marginal covariance for the study $s$ is: \(\Sigma_s=\Phi^*T_s{\Phi^*}^\top + \Psi_s = \Phi^*P{\Phi^*}^\top +\Phi^*R_s{\Phi^*}^\top+\Psi_s\). The common covariance component is extracted by \(\Phi^*P{\Phi^*}^\top\), and the study-specific covariance is \(\Phi^*R_s{\Phi^*}^\top\).

\paragraph{Priors}
Tetris places MGPS priors on $\Phi^*$ to encourage shrinkage and reduce overfitting. For the factor indicator matrix, i.e., $\mathcal{T}$, it adopts an Indian Buffet Process (IBP) prior \cite{knowles2007infinite}, a nonparametric prior well-suited for modeling latent binary matrices with potentially infinite columns. The IBP is controlled by hyperparameters $\alpha_{\mathcal{T}}$ and $\beta_{\mathcal{T}}$, which regulate the expected number of active factors and their distribution across studies. By default, $\alpha_{\mathcal{T}} = 1.25 \times S$ and $\beta_{\mathcal{T}} = 1$.

Each residual variance $\psi_{ps}$ again follows an inverse-gamma prior, with hyperparameters set as in BMSFA.

 \paragraph{Estimation}
 Tetris uses a Metropolis-within-Gibbs sampler\cite{gelman1995bayesian}. The factor allocation matrix $\mathcal{T}$ is updated using a Metropolis-Hastings step, while all other parameters (including $\Phi^*$, $\Psi_s$, and latent factors) are sampled using the Gibbs sampler. 

 Once the MCMC chains of $\mathcal{T}$ are obtained, the point estimate of $\mathcal{T}$ is selected by identifying the matrix that lies in the mode of the posterior—specifically, the configuration with the highest local density under a neighborhood metric.   Conditional on the selected $\mathcal{T}$, the model is re-fit using standard Gibbs sampling to obtain aligned posterior samples of factors and loadings with consistent dimensions. 

The number of factors is computed via the estimated $\mathcal{T}$ matrix. The number of common factors $K$ is the number of columns containing all $1$s in $\mathcal{T}$, while the number of study-specific factors for study $s$ refers to the number of $1$s in the $s$th row in $\mathcal{T}$, subtracting by $K$.
 
In summary, Table \ref{tab:summary} summarizes the method we discussed above and their priors.

\newcolumntype{L}[1]{>{\raggedright\arraybackslash}p{#1}}
\begin{longtable}{L{1.2cm} L{4.4cm} L{4.8cm} L{5.6cm}}
    \caption{\it Summary of the Bayesian integrative factor analysis models considered in this study, including their prior assumptions and structural specifications.}\label{tab:summary} \\
    \toprule
    \textbf{Model} & \textbf{Model formula} & \textbf{Covariance decomposition} & \textbf{Priors} \\
    \midrule
    \endfirsthead

    \caption[]{(continued)} \\
    \toprule
    \textbf{Model} & \textbf{Model formula} & \textbf{Covariance decomposition} & \textbf{Priors} \\
    \midrule
    \endhead

    \midrule
    \endfoot

    \bottomrule
    \endlastfoot
    
    \textbf{Stack FA} &
    \(\begin{aligned}
        \mathbf{y}_{is} &= \Phi \mathbf{f}_{is} + \boldsymbol{\epsilon}_{is} \\
        \mathbf{f}_{is} &\sim \text{MVN}(0, I_K) \\
        \boldsymbol{\epsilon}_{is} &\sim \text{MVN}(0, \Psi)\\
        \Psi &= \text{diag}(\psi_{1}, \psi_{2}, \ldots, \psi_{P})
    \end{aligned}\) &
    \( \begin{aligned}
        &\text{Marginal covariance:} \\
        & \quad \Sigma_s = \Phi \Phi^\top + \Psi \\
        &\text{Common covariance:} \\
        & \quad\Sigma_{\Phi} = \Phi \Phi^\top
    \end{aligned} \) &
    \(\begin{aligned}
        \text{For } & \Phi: \\
        &\phi_{pk}|\omega_{pk}, \theta \sim \mathcal{N}(0, \omega_{pk}^{-1} \theta_k^{-1}) \\
        &\omega_{pk} \sim \text{Gam}(\kappa/2, \kappa/2) \\
        &\theta_k = \prod_{l=1}^{k} \delta_l, \quad \delta_1 \sim \text{Gam}(a_1, 1), \\ 
        &\delta_l \sim \text{Gam}(a_2, 1), \text{ for }l > 1 \\
        \text{For } & \Psi: \\
        &\psi_{p} \sim \text{IG}(a_\psi, b_\psi)
    \end{aligned}\) \\
    
    \rowcolor{gray!20} \textbf{Ind FA} &
    \(\begin{aligned}
        \mathbf{y}_{is} &= \Lambda_s \mathbf{l}_{is} + \boldsymbol{\epsilon}_{is} \\
        \mathbf{l}_{is} &\sim \text{MVN}(0, I_{J_s}) \\
        \boldsymbol{\epsilon}_{is} &\sim \text{MVN}(0, \Psi_s)\\
        \Psi_s &= \text{diag}(\psi_{1s}, \psi_{2s}, \ldots, \psi_{Ps})
    \end{aligned}\) &
    \( \begin{aligned}
         & \text{Marginal covariance:}\\
         & \quad \Sigma_s = \Lambda_s \Lambda_s^\top + \Psi_s \\
        & \text{Study-specific covariance:} \\
        &\quad \Sigma_{\Phi} = \Lambda_s \Lambda_s^\top
    \end{aligned} \) &
    \(\begin{aligned}
        \text{For } & \Lambda_s: \\
        &\lambda_{pm}|\omega_{pm}, \theta_m \sim \mathcal{N}(0, \omega_{pm}^{-1} \theta_m^{-1}) \\
        &\omega_{pm} \sim \text{Gam}(\kappa^s/2, \kappa^s/2) \\
        &\theta_m = \prod_{l=1}^{m} \delta_l, \quad \delta_1 \sim \text{Gam}(a_1^s, 1), \\ 
        &\delta_l \sim \text{Gam}(a_2^s, 1), \text{ for }l > 1 \\
        \text{For } & \Psi_s: \\
        &\psi_{ps} \sim \text{IG}(a_\psi, b_\psi)
    \end{aligned}\) \\

    \textbf{PFA} &
    \(\begin{aligned}
        Q_s & \mathbf{y}_{is} = \Phi \mathbf{f}_{is} + \boldsymbol{\epsilon}_{is} \\
        Q_s &\sim \text{MN}_{P \times P}(I_P, \alpha_Q I_P, \alpha_Q I_P), \\
        \quad & Q_1 = I_P \\
        \mathbf{f}_{is} &\sim \text{MVN}(0, V) \\
        V &= \text{diag}(\nu_1, \nu_2, \ldots, \nu_K) \\
        \boldsymbol{\epsilon}_{is} &\sim \text{MVN}(0, \Psi)\\
        \Psi &= \text{diag}(\psi_1, \psi_2, \ldots, \psi_P)
    \end{aligned}\) &
    \( \begin{aligned}
        &\text{Marginal covariance:} \\
        &\quad \Sigma_s = Q_s^{-1}(\Phi V \Phi^\top + \Psi)({Q_s^{-1}})^\top \\
        &\text{Common covariance:} \\
        &\quad \Sigma_{\Phi} = \Phi V \Phi^\top + \Psi\\
        &\text{Study-specific covariance:} \\
        &\quad\Sigma_{\Lambda_s} = \Sigma_s - \Sigma_{\Phi}
    \end{aligned} \) &
    \(\begin{aligned}
        \text{For } & \Phi:\\
        &\phi_{pk}|\omega_{pk}, \theta \sim \mathcal{N}(0, \omega_{pk}^{-1} \theta_k^{-1}) \\
        &\omega_{pk} \sim \text{Gam}(\kappa/2, \kappa/2) \\
        &\theta_k = \prod_{l=1}^{k} \delta_l, \quad \delta_1 \sim \text{Gam}(a_1, 1), \\ 
        &\delta_l \sim \text{Gam}(a_2, 1), \quad \text{for }l > 1 \\
        \text{For }&V \text{ and } \Psi:\\
        & v_k \sim \text{IG}(a_\nu, b_\nu)\\
        & \psi_p \sim \text{IG}(a_\psi, b_\psi)
    \end{aligned}\) \\
    
     \rowcolor{gray!20} \textbf{MOM-SS} &
    \(\begin{aligned}
        \mathbf{y}_{is} &= \boldsymbol{\alpha}_s + \boldsymbol{\beta} \mathbf{x}_{is} + \Phi \mathbf{f}_{is} + \boldsymbol{\epsilon}_{is} \\
        \mathbf{f}_{is} &\sim \text{MVN}(0, I_K) \\
        \boldsymbol{\epsilon}_{is} &\sim \text{MVN}(0, \Psi_s)\\
        \Psi_s &= \text{diag}(\psi_{1s}, \psi_{2s}, \ldots, \psi_{Ps})
    \end{aligned}\) &
    \( \begin{aligned}
         &\text{Marginal covariance:}\\
         &\quad \Sigma_s = \Phi \Phi^\top + \Psi_s \\
        &\text{Common covariance:} \\
        &\quad \Sigma_{\Phi} = \Phi \Phi^\top
    \end{aligned} \) &
    \(\begin{aligned}
        \text{For }&\Phi:\\
        &\begin{aligned}
        \phi_{pk}|\gamma_{pk}, \tau_0, \tau_1
        =& (1 - \gamma_{pk}) f_{\text{spike}} \\ 
           &+ \gamma_{pk} f_{\text{slab}} 
        \end{aligned}\\
        &\gamma_{pk} \sim \text{Bernoulli}(\zeta_k) \\
        &\zeta_k \sim \text{Beta}(\frac{a_\zeta}{k}, b_\zeta) \\
        &f_{\text{spike}} = \mathcal{N}(0, \tau_0) \\
        &f_{\text{slab}} = (\phi_{pk}^2 / \tau_1) \mathcal{N}(0, \tau_1) \\
        \text{For }& \alpha_{ps} \text{ and } \beta_p:\\
        &\alpha_{ps}, \beta_p \sim \mathcal{N}(0, \sigma^2_\text{reg}I) \\
        \text{For }&\Psi_s:\\
        &\psi_{ps} \sim \text{IG}(a_\psi, b_\psi)
    \end{aligned}\) \\
    
    \textbf{SUFA} &
    \(\begin{aligned}
        \mathbf{y}_{is} &= \Phi \mathbf{f}_{is} + \Phi A_s\mathbf{l}_{is} + \boldsymbol{\epsilon}_{is} \\
        \mathbf{f}_{is} &\sim \text{MVN}(0, I_K) \\
        \mathbf{l}_{is} &\sim \text{MVN}(0, I_{J_s}) \\
        \boldsymbol{\epsilon}_{is} &\sim \text{MVN}(0, \Psi)\\
        \Psi &= \text{diag}(\psi_1, \psi_2, \ldots, \psi_P)
    \end{aligned}\) &
    \( \begin{aligned}
        &\text{Marginal covariance:} \\
        &\quad \Sigma_s = \Phi\Phi^\top + \Phi A_sA_s^\top \Phi^\top + \Psi\\
        &\text{Common covariance:} \\
        &\quad \Sigma_{\Phi} = \Phi\Phi^\top + \Psi\\
        &\text{Study-specific covariance:} \\
        &\quad\Sigma_{\Lambda_s} = \Phi A_sA_s^\top \Phi^\top
    \end{aligned} \) &
    \(\begin{aligned}
        \text{For }&\Phi:\\
        &\text{vec}(\Phi)\sim \text{DL}(a_{DL}) \\
        \text{For }&A_s:\\
        &a_{pjs} \sim \text{N}(0,\sigma^2_{A}) \\
        \text{For }&\Psi:\\
        &\psi_p \sim \text{Log-Normal}(\mu_{\psi}, \sigma_{\psi}^2) \\
    \end{aligned}\) \\

     \rowcolor{gray!20} \textbf{BMSFA} &
    \(\begin{aligned}
        \mathbf{y}_{is} &= \Phi \mathbf{f}_{is} + \Lambda_s \mathbf{l}_{is} + \boldsymbol{\epsilon}_{is} \\
        \mathbf{f}_{is} &\sim \text{MVN}(0, I_K) \\
        \mathbf{l}_{is} &\sim \text{MVN}(0, I_{J_s}) \\
        \boldsymbol{\epsilon}_{is} &\sim \text{MVN}(0, \Psi_s)\\
        \Psi_s &= \text{diag}(\psi_{1s}, \psi_{2s}, \ldots, \psi_{Ps})
    \end{aligned}\) &
    \( \begin{aligned}
        &\text{Marginal covariance:} \\
        &\quad \Sigma_s = \Phi \Phi^\top + \Lambda_s \Lambda_s^\top + \Psi_s\\
         & \text{Common covariance:}\\
         &\quad\Sigma_{\Phi} = \Phi \Phi^\top\\
         &\text{Study-specific covariance:}\\
         &\quad\Sigma_{\Lambda_s} = \Lambda_s \Lambda_s^\top
    \end{aligned} \) &
    \(\begin{aligned}
        \text{For } & \Phi:\\
        &\phi_{pk}|\omega_{pk}, \theta \sim \mathcal{N}(0, \omega_{pk}^{-1} \theta_k^{-1}) \\
        &\omega_{pk} \sim \text{Gam}(\kappa/2, \kappa/2) \\
        &\theta_k = \prod_{l=1}^{k} \delta_l, \quad \delta_1 \sim \text{Gam}(a_1, 1), \\ 
        &\delta_l \sim \text{Gam}(a_2, 1), \quad \text{for }l > 1 \\
        \text{For }&\Lambda_s:\\
        &\lambda_{pm}|\omega_{pm}, \theta_{m} \sim \mathcal{N}(0, \omega_{pm}^{-1} \theta_m^{-1}) \\
        &\omega_{pm} \sim \text{Gam}(\kappa_s/2, \kappa_s/2) \\
        &\theta_m = \prod_{l=1}^{m} \delta_l, \quad \delta_1 \sim \text{Gam}(a_1^s, 1), \\ &\delta_l \sim \text{Gam}(a_2^s, 1), \quad \text{for }l > 1 \\
        \text{For }&\Psi_s:\\
        &\psi_{ps} \sim \text{IG}(a_\psi, b_\psi)
    \end{aligned}\) \\
    
    \textbf{Tetris} &
    \(\begin{aligned}
        \mathbf{y}_{is} &= \Phi^* T_s\mathbf{f}_{is} + \boldsymbol{\epsilon}_{is}, \\
        T_s&=\text{diag}(t_{1s},\cdots, t_{Ks})\\
        &\text{ with } t_{ks} \in \{0, 1\}\\
        \mathbf{f}_{is} &\sim \text{MVN}(0, I_K) \\
        \boldsymbol{\epsilon}_{is} &\sim \text{MVN}(0, \Psi_s)\\
        \Psi_s &= \text{diag}(\psi_{1s}, \psi_{2s}, \ldots, \psi_{Ps})
    \end{aligned}\) &
    \( \begin{aligned}
        &\text{Marginal covariance:} \\
        &\quad \Sigma_s = \Phi^* T_s  (\Phi^*)^\top + \Psi_s \\
        &\text{Common covariance:} \\
        &\quad \Sigma_{\Phi} =  \Phi^* P(\Phi^*)^\top\\
        &\text{Study-specific covariance:} \\
        &\quad\Sigma_{\Lambda_s} =  \Phi^* R_s(\Phi^*)^\top\\
        &\text{where, }  T_s = P+R_s, \\
        &\text{with } P=\text{diag}(\mathbbm{1}_{1},\cdots, \mathbbm{1}_{K^*}),\\
        &\quad\mathbbm{1}_{k}= 
        \begin{cases}
            1 &\text{if } t_{ks} =1 \text{ for all } s.\\
            0 & \text{otherwise}.
        \end{cases}
    \end{aligned} \) &
    \(\begin{aligned}
        \text{For }&  \Phi^*:\\
        &\phi^*_{pk}|\omega_{pk}, \theta \sim \mathcal{N}(0, \omega_{pk}^{-1} \theta_k^{-1}) \\
        &\omega_{pk} \sim \text{Gam}(\kappa/2, \kappa/2) \\
        &\theta_k = \prod_{l=1}^{k} \delta_l, \quad \delta_1 \sim \text{Gam}(a_1, 1), \\ 
        &\delta_l \sim \text{Gam}(a_2, 1), \quad \text{for }l > 1 \\
        \text{For }&T_s:\\
        &\mathcal{T}\sim \text{Indian Buffet Process}(\alpha_{\mathcal{T}}, \beta_{\mathcal{T}}),\\
        &\text{where } s\text{th row of } \mathcal{T}\text{ contains the}\\
        &\text{ diagonal entries of }T_s. \\
        \text{For }&\Psi_s:\\
        &\psi_{ps} \sim \text{IG}(a_{\psi}, b_{\psi}) \\
    \end{aligned}\) \\
\end{longtable}

\subsection{Identifiability Issues and Post-processing}\label{sec:Identifiability}

A central challenge in factor models is rotational invariance, whereby the model remains unchanged under orthogonal transformations of the latent space. Specifically, if $\Phi$ denotes the factor loading matrix and $\mathbf{F}$ the matrix of factor scores, then for any orthogonal matrix $\Gamma$, the transformation $\Phi' = \Phi \Gamma$ and $\mathbf{F}' = \mathbf{F} \Gamma^\top$ yields the same fitted values: $\mathbf{Y} = \mathbf{F} \Phi^\top = \mathbf{F}' {\Phi'}^\top$. Two special cases of such transformations are column permutation (label switching) and sign flipping (multiplying a column by $-1$).
In MCMC-based approaches, posterior may differ by such transformations,  leading to non-identifiable or misleading posterior summaries (e.g., marginal means).  Similarly, in EM-based algorithms, multiple local optima can yield distinct but statistically equivalent solutions.

To address these issues, two broad strategies are commonly employed:
\begin{enumerate}
  \item Impose parameters' constraints, such as assuming heteroscedastic factors or lower-triangular loading matrices to break the rotational invariance.
  \item Post-processing the estimated loadings or posterior samples using rotation and alignment techniques, such as orthogonal Procrustes (OP)\cite{assmann2016bayesian}, etc.
\end{enumerate}

Below, we briefly describe how each model in this tutorial (Stack FA, Ind FA, PFA, MOM-SS, SUFA, BMSFA, Tetris) handles identifiability:

\paragraph{Stack FA \& Ind FA}
Both Stack FA and Ind FA are typically estimated via a Gibbs sampling. 
 To address identifiability {\it post hoc}, we adopt the Orthogonal Procrustes (OP) alignment \cite{assmann2016bayesian}, which aligns posterior samples to a reference loading matrix by minimizing the  differences between the posterior draws and a reference loading matrix . This results in consistent orientation and factor labeling across MCMC draws.
Alternatively, one can use Spectral Decomposition (SD). After estimating the covariance $\widehat{\Sigma}$, SD decomposes it as $\widehat{\Sigma} = UNU^\top$, where $U$ contains the eigenvectors and $N$ the eigenvalues. Retaining the top $K^*$ eigenvectors and scaling by the square roots of their eigenvalues, the loadings matrix is reconstructed as $\widehat{\Phi}=U_{K^*}\,N_{K^*}^{1/2}$. 
 This is similar to standard PCA-based factor analysis \cite{darton1980rotation}.
A third option is Varimax rotation \cite{kaiser1958varimax}, which  seeks an orthogonal transformation that maximizes the variance of squared loadings within each factor. This approach enhances interpretability by encouraging sparsity and grouping high-loading variables. More recent advancements addressing identifiability can be found in \cite{papastamoulis2022identifiability}.

\paragraph{PFA}
Perturbed Factor Analysis (PFA) mitigates rotational ambiguity by assigning each latent factor its own variance (heteroscedastic factors). Specifically, post-processing alignment is applied based on an extension of the Procrustes method, as described in Roy et al.\cite{roy2021bayesian} to provide uncertainty quantification (UQ)\cite{roy2021bayesian}. This approach yields consistent factor labeling across posterior draws.

\paragraph{MOM-SS}
MOM-SS model \cite{Avalos2022HLDI} applies a Varimax rotation during initialization to improve convergence and avoid spurious local maxima. Furthermore, the non-local spike-and-slab prior \cite{johnson2012bayesian} induces sparsity, shrinking many loadings  toward zero, which helps reduce the space of equivalent solutions and mitigates label-switching.

\paragraph{SUFA}
SUFA \cite{chandra2024inferring} addresses identifiability through a combination of design and post-processing. First, it constrains the model via the inequality $\sum_{s=1}^S J_s \le K$ and uses continuous priors on $A_s$ to avoid identifiability issues. Second, it applies a rotate-and-align procedure based on Varimax, as proposed in \cite{poworoznek2021efficiently}, to ensure a consistent orientation of the factors across MCMC draws.

\paragraph{BMSFA}
BMSFA \cite{de2021bayesian} uses the OP method to resolve rotational invariance in the posterior samples, consistent with Stack FA and Ind FA.

\paragraph{Tetris}
Tetris \cite{grabski2023bayesian} handles identifiability by selecting the posterior draw of $\mathcal{T}$ (the factor activation matrix) that lies in the mode of the MCMC samples. Factor loadings are aligned across posterior draws by minimizing a mean squared error (MSE)-like loss between each draw’s implied covariance and the estimated marginal covariance. This effectively enforces consistent labeling and orientation across iterations.

\medskip In summary, all models discussed here must address the fundamental challenge of orthogonal non-identifiability. Most achieve this via {\it post hoc} alignment,
or via structural constraints or sparsity-inducing priors. Additional post-processing, such as factor reordering by explained variance or total absolute loadings \cite{griffiths2011indian}, can be applied to highlight the most important factors.

\section{Simulation}\label{sec:3}

\subsection{Simulation settings}

We assess the performance of the seven previously described Bayesian integrative factor models in recovering factor loadings and estimating the number of latent factors across multi-study datasets with diverse latent structures.

The models are implemented as follows:

\begin{itemize}
    \item StackFA, IndFA, and BMSFA using the \texttt{MSFA} R package (v0.86);
    \item PFA using the script "FBPFA-PFA with fixed latent dim.R" from the GitHub repository \href{https://github.com/royarkaprava/Perturbed-factor-model/tree/master}{https://github.com/royarkaprava/Perturbed-factor-model};
\item MOM-SS using the \texttt{BFR.BE} R package (v0.1.0);
\item SUFA using the \texttt{SUFA} R package (v2.2.0);
\item Tetris using the script from \href{https://github.com/igrabski/tetris/tree/main}{https://github.com/igrabski/tetris/tree/main}.
\end{itemize}

Furthermore, we evaluate two versions of SUFA: \texttt{SUFA\_fixJs}, which requires specifying the number of study-specific factors, and \texttt{SUFA}, which infer this quantity automatically. Similarly, we evaluate two variants of Tetris: \texttt{Tetris}, the standard approach recommended by the author, and \texttt{Tetris\_fixT}, which pre-specifies the number of factors to improve scalability.

We consider five different scenarios. The first three are based on the data-generating processes from PFA, MOM-SS, and SUFA, respectively. The fourth and fifth are more general and are designed to closely mimic the two case studies presented in Section \ref{sec:4}; their data-generating process is based on Tetris, which is more complex and involves more parameters than all the other methods.

For each scenario, we generated 50 collections of datasets. The data generation code is adapted from publicly available repositories of PFA, MOM-SS, and BMSFA. In all scenarios, the data are centered (but not scaled) before model fitting, except for MOM-SS, which incorporates study-specific intercepts to account for uncentered data. For MCMC-based methods, we run 10,000 iterations with a burn-in of 8,000. Post-processing is carried out following the procedures described in Sections \ref{sec:Identifiability} and \ref{sec:post-nutrition}.

\textbf{Scenario 1: based on PFA}

In this scenario, data are generated from four studies ($S=4$), each with 100 samples ($N_s=100$) , for a total of 400 observations, and the number of common factors $K=4$. Each sample includes 40 observed variables ($P=40$). The common factor loading matrix $\Phi$ is generated  with 40\% zero entries, and the remaining values are drawn from U$(0.6, 1)$ and assigned a negative sign with 0.5 probability. The residual covariance $\Psi$ is set to be the same across all studies, with diagonal elements drawn from U$(0,1)$.  After generating $\mathbf{Y}_s = \mathbf{F}_s\Phi^\top + \mathbf{E}_s$, we apply a perturbation matrix $Q_s$ with perturbation level $\alpha_Q = 0.01$ and compute $\mathbf{Y}_s^{\text{perturbed}} = Q_s \mathbf{Y}_s$.

The ground truth quantities used for evaluation include the common loadings $\Phi$, the common covariance matrix $\Sigma_\Phi = \Phi\Phi^\top + \Psi$, the study-specific covariance matrices $\Sigma_{\Lambda_s} = \Sigma_s - \Sigma_\Phi$, and the marginal covariances $\Sigma_s = Q_s \Sigma_\Phi Q_s^\top$. 

There is no explicit definition of study-specific loadings in the PFA model. While a potential derivation is $\Lambda_s = (Q_s^{-1} - I_P)\Phi$, this yields $\Lambda_1 = \mathbf{0}$ for $Q_1 = I_P$, making it inconsistent for comparative purposes.

Each method is fit with the true number of common factors $K = 4$, with the exception of IndFA, which does not include the common factors and thus we set the number of study-specific factors $J_s$ to 4 for each study. Tetris also has a version that does not require pre-specification of $K$. For those models that require $J_s$, we set $J_s = 1$ for SUFA\_fixJs and BMSFA. For Tetris\_fixT, we predefine the matrix for factor structure $\mathcal{T}$ to include 4 common factors and 1 study-specific factor for each study.

\textbf{Scenario 2: based on MOM-SS}

In this scenario, the data are again generated from four studies ($S = 4$), each consisting of 100 samples, for a total of 400 observations. Each sample includes 40 observed variables ($P = 40$), and the data-generating process incorporates common factors ($K = 4$), study-specific intercepts ($\boldsymbol{\alpha}_s$), two observed covariates ($Q = 2$), and noise terms drawn from a multivariate normal distribution with study-specific idiosyncratic covariance matrices. The common loading matrix $\Phi$ is constructed to be sparse, with 40\% of entries set to zero. The remaining entries are drawn uniformly from the interval $[0.6, 1]$ and randomly assigned negative signs with probability 0.5.

The quantities used for evaluation in this scenario include the common loading matrix $\Phi$, the common covariance matrix $\Sigma_\Phi = \Phi\Phi^\top$, and the total covariances in each study, $\Sigma_s = \Phi\Phi^\top + \Psi_s$. Since MOM-SS does not include study-specific factor loadings in its generative process, there are no true $\Lambda_s$ or $\Sigma_{\Lambda_s}$ in this setting.

Each model is fit using the same factor specifications as in Scenario 1.

\textbf{Scenario 3: based on SUFA}

In this scenario, data are generated from four studies ($S = 4$), each containing 100 samples, for a total of 400 observations. Each sample includes 40 observed variables ($P = 40$), generated from a multivariate normal distribution with covariance structure $\Sigma_s = \Phi\Phi^\top + \Phi A_s A_s^\top \Phi^\top + \Psi$. The common loading matrix $\Phi$ is generated with $K = 4$ factors and follows the same construction as in the previous scenarios, with 40\% sparsity and nonzero entries drawn from a uniform distribution on $[0.6, 1]$, with random sign flipping. The study-specific factor structure is introduced via the matrices $A_s$, which are drawn from a multivariate normal distribution centered at zero with standard deviation 0.4. 

The true quantities used for evaluation are the common loadings $\Phi$, the common covariance $\Sigma_\Phi = \Phi\Phi^\top + \Psi$, the study-specific loadings $\Lambda_s = \Phi A_s$, the study-specific covariances $\Sigma_{\Lambda_s} = \Phi A_s A_s^\top \Phi^\top$, and the marginal covariances $\Sigma_s$.

For model fitting, we use the true values $K = 4$ and $J_s = 1$ wherever the method requires them. In the case of IndFA, , we set $J_s = 5$ for each study.

\textbf{Scenario 4: mimic nutrition data and based on Tetris}

In this scenario, the data are designed to closely mimic the  real case nutrition dataset analyzed in Section~\ref{sec:4}. Data are generated from twelve studies ($S = 12$), with varying sample sizes per study: $N_s = (1362, 217, 417, 1012, 2241, 205, 2403, 3775, 1790, 761, 373, 465)$, resulting in a total sample size of 16,021. Each study includes 42 observed variables ($P = 42$), and 12 covariates $X$. 

For each individual, the observed response vector $\mathbf{y}_{is}$ is generated from a multivariate normal distribution with covariance $\Sigma_s = \Phi^* T_s (\Phi^*)^\top + \Psi_s$, following the Tetris model. We also added the covariates and random residuals. The combinatorial loading matrix $\Phi^*$ is generated as in the previous scenarios, with sparsity and uniformly distributed non-zero entries. The latent structure encoded in $T_s$ assigns $K = 4$ common factors, $J_s = 1$ study-specific factor for each study, and 7 partially shared factors distributed across studies.

The ground truth includes the common loadings $\Phi$, the study-specific loadings $\Lambda_s$, the common covariance $\Sigma_\Phi$, the study-specific covariances $\Sigma_{\Lambda_s}$, and the marginal covariances $\Sigma_s$, as summarized in Table~\ref{tab:summary}.

When fitting the models, we specify $K = 4$ for all  methods. For SUFA\_fixJs and BMSFA, we set $J_s = 1$ for each study, acknowledging that this may slightly violate the identifiability conditions for SUFA. For IndFA, we specify $J_s = 5$ to provide flexibility for modeling potentially shared factors. Tetris\_fixT is fit using the exact structure $\mathcal{T}$ employed during data generation, ensuring optimal alignment with the truth.

\textbf{Scenario 5: mimic gene expression data and based on Tetris}

In this final scenario, the data are constructed to closely mimic the real case study of gene expression used in Section~\ref{sec:4}, i.e., curatedOvarianData\cite{ganzfried2013curatedovariandata}. The simulation includes four studies ($S = 4$), with study-specific sample sizes given by $N_s = (157, 195, 285, 117)$. Each sample contains high-dimensional $P = 1060$ gene expression variables. The data are generated under the Tetris model, incorporating $K = 15$ common factors, $J_s = 2$ study-specific factors for each study, and 3 partially shared factors.

The loading matrix $\Phi^*$ is generated as in previous scenarios, but with increased sparsity: 80\% of its entries are set to zero. The non-zero elements are drawn from a uniform distribution over $[0.6, 1]$ and are randomly signed. The covariance matrices $\Sigma_s$ are constructed according to the Tetris model: $\Sigma_s = \Phi^* T_s (\Phi^*)^\top + \Psi_s$. The true quantities for evaluation are the common loadings $\Phi$, study-specific loadings $\Lambda_s$, common covariance $\Sigma_\Phi$, study-specific covariances $\Sigma_{\Lambda_s}$, and marginal covariances $\Sigma_s$, as detailed in Table~\ref{tab:summary}.

For model fitting, we set $K = 15$ for all methods, and $J_s = 2$ for SUFA\_fixJs and BMSFA. For IndFA, we increase flexibility by specifying $J_s = 17$ for each study, to account for both study-specific and potentially shared variation. Tetris\_fixT is applied using the exact $\mathcal{T}$ matrix of data generation. Due to high computational demands, SUFA and SUFA\_fixJs are only repeated 30 times in this scenario.

\textbf{Over-specifying the numbers of factors}

To assess the ability of each method to recover the correct number of common ($K$) and study-specific ($J_s$) factors, we repeat the simulations from all five scenarios, this time over-specifying the number of factors when fitting the models. Tetris, which internally determines its latent structure, is evaluated using the same results as in the correctly specified setting.

For Scenarios 1–3, we set $K = 6$ in StackFA, PFA, MOM-SS, SUFA, and BMSFA, and $J_s = 2$ in BMSFA and $J_s = 6$ in IndFA, for all studies ($s = 1, \dots, 4$). For Scenario 4, we again specify $K = 6$ for StackFA, PFA, MOM-SS, SUFA, and BMSFA, $J_s = 2$ for BMSFA, and $J_s = 6$ for IndFA across all twelve studies. For Scenario 5, we set $K = 20$ in StackFA, MOM-SS, SUFA, and BMSFA, $J_s = 4$ for BMSFA, and $J_s = 20$ for IndFA, for $s = 1, \dots, 4$. In this setting, SUFA\_fixJs and Tetris\_fixT are not run due to their reliance on fixed, known factor configurations.

For models that estimate the number of factors internally,namely MOM-SS, SUFA,  and Tetris, we extract the inferred $K$ and $J_s$ directly from the dimensions of their estimated loading matrices. For PFA, which uses an adaptive truncation strategy within the MCMC procedure, we estimate $K$ by computing the mode of the number of active columns across posterior samples. In contrast, StackFA, IndFA, and BMSFA do not estimate the number of factors directly. Instead, we apply eigenvalue decomposition (EVD) to their estimated covariance matrices and determine the number of factors as those explaining more than 5\% of the total variance.

\subsection{Simulation results: factor loadings and computational efficiency}

\begin{figure}[!ht]
    \centering
    \includegraphics[width=0.8\linewidth]{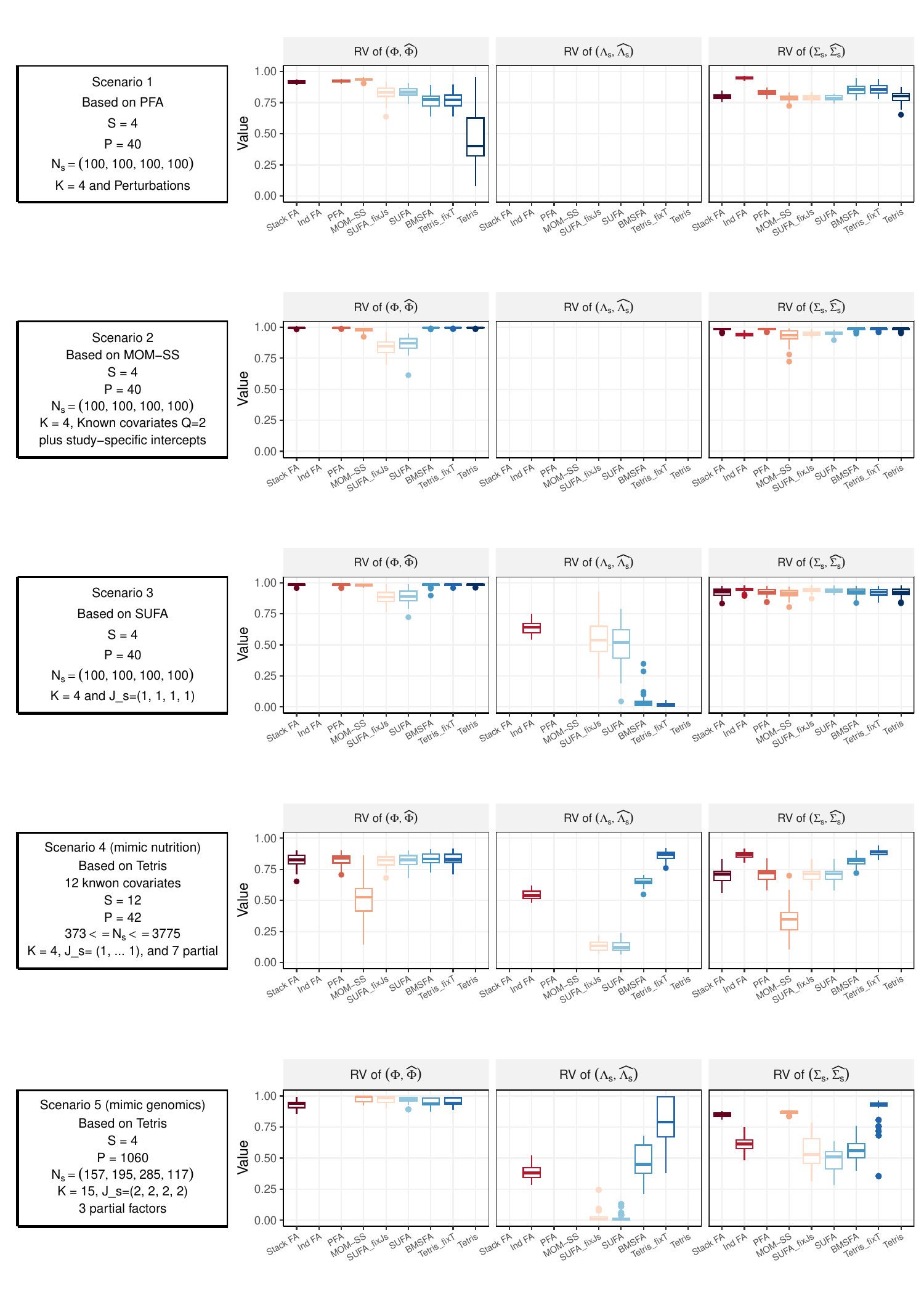}

    \caption{\it  Measuring accuracy of different Bayesian integrative factor models in estimating factor loadings and marginal covariance matrices, across five simulation scenarios with correctly specified numbers of factors. The first column lists out the settings of a scenario, the second column shows the RV  coefficient between estimated and  true  common loadings ($\Phi$), the third column shows the RV or study-specific loadings ($\Lambda_s$), and the third column shows the RV for the marginal covariance matrices ($\Sigma_s$). In Scenarios 1 and 2, the study-specific loadings $\Lambda_s$ are not part of the generative model (PFA and MOM-SS, respectively), so the corresponding panels are blank. StackFA and IndFA estimate only common or study-specific loadings, respectively, and thus may have missing values in some panels. PFA fails to run in Scenario 5 due to memory and time constraints.  Tetris runs for over 24 hours in Scenario 4 and 5. Thus we do not report results for those models in those scenarios.}
    \label{fig:box_acc}
\end{figure}

\begin{figure}[!ht]
    \centering
    \includegraphics[width=0.8\linewidth]{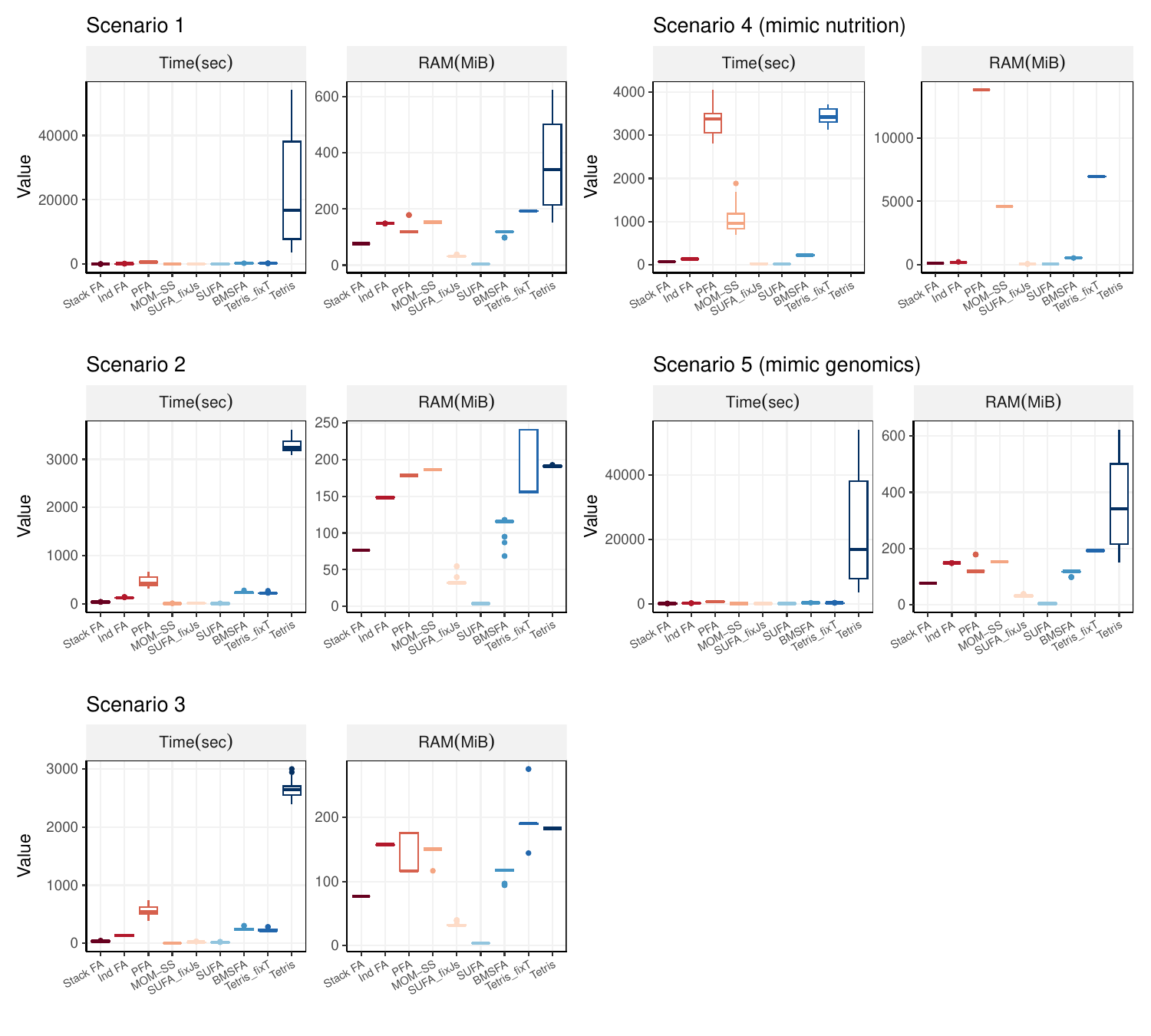}
    \caption{\it Computational efficiency of Bayesian integrative factor models, evaluated under correctly specified numbers of factors across the five simulation scenarios. The runtime (in seconds) and peak memory usage (in MiB) are recorded for each model.}
    \label{fig:box_eff}
\end{figure}

We evaluate the performance of each model in recovering the common loadings $\Phi$, study-specific loadings $\Lambda_s$, common covariances $\Sigma_\Phi$, study-specific covariances $\Sigma_{\Lambda_s}$, and marginal covariances $\Sigma_s$ by comparing them to the ground truth. Two metrics are used: the RV coefficient~\cite{robert1976unifying} and the Frobenius norm (FN)~\cite{horn2012matrix} (see Supplementary for full FN definitions and results).

The RV coefficient, implemented via the \texttt{RV} function in the R package \texttt{MatrixCorrelation}~\cite{MatrixCorrelation2018}, measures matrix similarity and is defined as:

\begin{equation}\label{eq:RV1}
    \text{RV}(X, Y)
    =\begin{aligned}
        \frac{\text{trace}\{XX^\top YY^\top\}}
        {\sqrt{(\text{trace}\{XX^\top XX^\top\})\times(\text{trace}\{YY^\top YY^\top\})}}
    \end{aligned}
\end{equation}

It ranges from 0 (no similarity) to 1 (perfect similarity). The Frobenius norm quantifies the element-wise difference between two matrices; values closer to 0 indicate better estimation accuracy.  For study-specific quantities, RV and FN scores are averaged across studies.

Computational performance is assessed by recording runtime (in seconds) and peak memory usage (in MiB).

Figures~\ref{fig:box_acc} and~\ref{fig:box_eff} present the results for each scenario, assuming the true number of factors is known. In Scenario 1 (PFA-generated data), most models accurately recover the common loadings and marginal covariances, although Tetris performs less well and takes substantially longer due to the difficulty in identifying small perturbations. In Scenario 2 (MOM-SS), all models perform well despite the presence of known covariates, since these effects are removed during data centering or captured on the study-specific effects.  
Scenarios 1 and 2 do not include the generation of study-specific loadings $\Lambda_s$, as the data-generating models (PFA and MOM-SS) do not incorporate the study-specific structure. As a result, $\Lambda_s$ is not used for precision comparisons in these settings; therefore, the corresponding panels for $\Lambda_s$ in Figure~\ref{fig:box_acc} are blank.

Scenario 3, generated from SUFA, features study-specific loadings constructed as $\Lambda_s = \Phi A_s$ with small perturbations. As a result, many models recover $\Phi$ and $\Sigma_s$ well, but fail to accurately estimate $\Lambda_s$; in particular, BMSFA and Tetris are less successful, with Tetris failing to detect any study-specific variation due to its weak signal.

In Scenario 4, which mimics the real nutrition data structure ($N_s \gg P$), Tetris\_fixT produces the most accurate estimates across all quantities due to perfect alignment with the generative model. BMSFA also performs well, but cannot account for partially shared factors. PFA achieves good accuracy but is computationally expensive, especially in memory usage. MOM-SS converges quickly, but struggles with precision when $N$ is large, probably due to premature termination of the EM algorithm  or convergence to a local maximum. SUFA underestimates study-specific loadings due to its model assumptions.

In Scenario 5, which mirrors high-dimensional genomic data ($P \gg N_s$), most models estimate the common loadings accurately. Tetris\_fixT again performs best due to its alignment with the underlying structure. SUFA remains comparable in accuracy to Scenario 4 but requires significantly more computational resources, as its complexity scales with $P$. PFA fails to complete within 24 hours and shows sensitivity to the dimensionality of the data.

Overall, model performance is scenario-dependent. StackFA and IndFA are limited by their structural assumptions, recovering only common or study-specific components but not both. These two methods used as benchmark also require lot of memory. PFA is accurate but slow, particularly with large $P$, due to the estimation of the $Q_s$ matrix of size $P \times P$ for each study. MOM-SS runs quickly using an EM algorithm, but loses accuracy in complex scenarios and tends to overestimate loadings. SUFA is efficient when $P$ is small, even for large $N_s$, and performs consistently across scenarios, although its speed advantage diminishes when $P$ is large. BMSFA is robust across scenarios but incurs longer runtimes due to Gibbs sampling. Tetris, while highly accurate, does not scale well to high-dimensional data. Its simplified variant, Tetris\_fixT, slightly reduces time and memory requirements while maintaining high estimation accuracy.

Models using MCMC algorithms, such as BMSFA, PFA, and Tetris, generally incur higher computational costs. MOM-SS is the fastest method overall due to its EM-based estimation, while SUFA remains efficient in scenarios with small to moderate $P$. Tetris shows the highest computational cost in Scenarios 4 and 5, exceeding 24 hours of runtime; PFA also fails to complete Scenario 5 within the time limit. Thus, results for Tetris in Scenarios 4–5 and for PFA in Scenario 5 are not available. Tetris\_fixT provides a faster and more memory-efficient alternative to Tetris while maintaining competitive accuracy.

\subsection{Simulation results: estimating the number of factors}

\begin{table}[h!]
\caption{\it Estimated number of common ($K$) and study-specific ($J_s$) factors across five simulation scenarios when models are fit with over-specified values. Values are reported as mean (standard deviation) over 50 simulation replicates (30 for SUFA in Scenario 5 due to high computational cost).}
\centering
\begin{tabular}{>{\raggedright\arraybackslash}p{2cm} p{2cm} p{11cm}}
  \hline
\rowcolor{gray!30}  \textbf{Model} & \textbf{Estimated $K$} & \textbf{Estimated $J_s$} \\ 
  \hline
  \multicolumn{3}{l}{\textbf{Scenario 1  True $K=4$, True $J_s = (0, 0, 0, 0)$}}  \\
\rowcolor{white}  Stack FA & 4.10(0.30) & -\\ 
    Ind FA & - & [4.00(0.00), 4.00(0.00), 4.00(0.00), 4.00(0.00)] \\ 
    PFA & 6.00(0.00) & -\\ 
   MOM-SS & 6.00(0.00) & -\\ 
   SUFA & 6.00(0.00) & [1.00(0.00), 1.00(0.00), 1.00(0.00), 1.00(0.00)] \\ 
  BMSFA & 3.06(0.47) & [2.00(0.00), 2.00(0.00), 2.00(0.00), 2.00(0.00)] \\ 
   Tetris & 1.38(1.14) & [4.66(2.21), 9.84(5.35), 10.16(5.04), 10.04(5.51)] \\ 
\hline
\rowcolor{gray!20} \multicolumn{3}{l}{\textbf{Scenario 2  True $K=4$, True $J_s = (0, 0, 0, 0)$}} \\
 \rowcolor{gray!20}    Stack FA & 4.00(0.00) & -\\ 
 \rowcolor{gray!20}   Ind FA & - & [4.00(0.00), 4.00(0.00), 4.00(0.00), 4.00(0.00)] \\ 
\rowcolor{gray!20}     PFA & 6.00(0.00) &- \\ 
 \rowcolor{gray!20}   MOM-SS & 6.00(0.00) & -\\ 
 \rowcolor{gray!20}   SUFA & 5.98(0.14) & [1.00(0.00), 1.00(0.00), 1.00(0.00), 1.00(0.00)] \\ 
 \rowcolor{gray!20}   BMSFA & 4.00(0.00) & [2.00(0.00), 2.00(0.00), 2.00(0.00), 2.00(0.00)] \\ 
\rowcolor{gray!20}    Tetris & 5.00(0.00) & [0.18(0.39), 0.12(0.33), 0.24(0.48), 0.18(0.44)] \\ 
\hline
 \multicolumn{3}{l}{\textbf{Scenario 3  True $K=4$, True $J_s = (1, 1, 1, 1)$}} \\
    Stack FA & 4.00(0.00) & -\\ 
    Ind FA & - & [4.00(0.00), 4.00(0.00), 4.00(0.00), 4.00(0.00)] \\ 
    PFA & 6.00(0.00) & -\\ 
   MOM-SS & 6.00(0.00) & -\\ 
   SUFA & 5.98(0.14) & [1.00(0.00), 1.00(0.00), 1.00(0.00), 1.00(0.00) \\ 
   BMSFA & 4.00(0.00) & [2.00(0.00), 2.00(0.00), 2.00(0.00), 2.00(0.00)] \\ 
   Tetris & 4.00(0.00) & [0.00(0.00), 0.00(0.00), 0.00(0.00), 0.00(0.00)] \\ 
   \hline
\rowcolor{gray!20} \multicolumn{3}{l}{\textbf{Scenario 4  True $K=4$, True $J_s = (1, 1, 1, 1, 1, 1, 1, 1, 1, 1, 1, 1)$}}\\
 \rowcolor{gray!20}   Stack FA & 6.00(0.00) & - \\ 
\rowcolor{gray!20}   Ind FA & - & [6.00(0.00), 6.00(0.00), 6.00(0.00), 6.00(0.00), 6.00(0.00), 6.00(0.00), 6.00(0.00), 6.00(0.00), 6.00(0.00), 6.00(0.00), 6.00(0.00), 6.00(0.00)] \\ 
 \rowcolor{gray!20}  PFA & 6.00(0.00) & - \\ 
 \rowcolor{gray!20}  MOM-SS & 6.00(0.00) & - \\ 
\rowcolor{gray!20}   SUFA & 6.00(0.00) & [1.00(0.00), 1.00(0.00), 1.00(0.00), 1.00(0.00), 1.00(0.00), 1.00(0.00), 1.00(0.00), 1.00(0.00), 1.00(0.00), 1.00(0.00), 1.00(0.00), 1.00(0.00)] \\ 
\rowcolor{gray!20}   BMSFA & 6.00(0.00) & [2.00(0.00), 2.00(0.00), 2.00(0.00), 2.00(0.00), 2.00(0.00), 2.00(0.00), 2.00(0.00), 2.00(0.00), 2.00(0.00), 2.00(0.00), 2.00(0.00), 2.00(0.00)] \\ 
\hline
\multicolumn{3}{l}{\textbf{Scenario 5  True $K=15$, True $J_s = (2, 2, 2, 2)$}} \\
    Stack FA & 10.90(1.04) & - \\ 
   Ind FA & - & [5.06(2.24), 7.10(1.58), 10.84(0.91), 2.34(1.57)] \\ 
  MOM-SS & 20.00(0.00) & - \\ 
  SUFA & 19.00(0.00) & [4.00(0.00), 4.00(0.00), 4.00(0.00), 4.00(0.00)] \\ 
   BMSFA & 11.76(0.85) & [2.00(0.00), 2.00(0.00), 1.70(0.46), 1.94(0.24)] \\ 
   \hline
\end{tabular}
\label{tab:unified_results}
\parbox{0.9\textwidth}{\footnotesize
\textit{Note}: Values are presented with mean(SD) over simulated 50 datasets.}
\end{table}

Table~\ref{tab:unified_results} summarizes the results. No method accurately recovers the correct number of factors across all scenarios. EVD-based approaches (used for StackFA, IndFA, and BMSFA) perform well in Scenarios 1–4, particularly for $J_s$ in IndFA, where partially shared factors are sometimes recovered. However, in Scenario 5, these methods tend to underestimate both $K$ and $J_s$, likely due to the high sparsity of the true loading matrix. PFA and MOM-SS consistently overestimate $K$, indicating overly permissive inclusion thresholds. SUFA also tends to overestimate $K$, while its estimates of $J_s$ remain stable across studies and scenarios. Tetris does not accurately recover $K$ and $J_s$ in Scenarios 1–3. Specifically, in Scenario 1, it fails to detect the small perturbation effects encoded in the data, leading to an underestimation of $K$ and overestimation of $J_s$, which also contributes to its long runtime. In Scenario 2, Tetris overestimates $K$ but gives the closest estimates for $J_s$. In Scenario 3, Tetris does not recover any study-specific factors, likely due to the small magnitude of $\Lambda_s$ generated via SUFA.

\section{Case study demonstration}\label{sec:4}

\subsection{Package installation and environment setup}

This tutorial is accompanied by an open-source R package and reproducible code repository, designed to guide researchers through the implementation and comparison of the Bayesian integrative factor models introduced in this paper. Our repository requires R version of 4.4.0 or greater, a dependency which is inherited from \texttt{Matrix} package version 1.7.

\textbf{Stack FA, Ind FA and BMSFA}

The Stack FA, Ind FA, and BMSFA models are implemented in the \texttt{MSFA} package, which is available on GitHub and can be installed using the \texttt{remotes} package:

\begin{lstlisting}[language=R]
install.packages("remotes")
remotes::install_github("rdevito/MSFA")
library(MSFA)
\end{lstlisting}

The function \texttt{sp\_fa()} is used to fit both the Stack FA model (using pooled data) and the Ind FA model (fitted separately to each study). The BMSFA model is fitted using \texttt{sp\_msfa()}.

\textbf{PFA}

The Perturbed Factor Analysis (PFA) model is not available as a standalone R package. Instead, the required R scripts must be obtained from the authors’ GitHub repository, i.e., \href{https://github.com/royarkaprava/Perturbed-factor-models}{https://github.com/royarkaprava/Perturbed-factor-models}.

Three files are needed to run the PFA model: \texttt{FBPFA-PFA.R}, \texttt{FBPFA-PFA} with fixed latent dim.R, and \texttt{PFA.cpp}. 
The \texttt{FBPFA-PFA.R} script provides the full Bayesian inference algorithm, where the common latent dimension is automatically set equal to the number of observed variables ($K = P$). The \texttt{FBPFA-PFA with fixed latent dim.R} script implements a variant requiring the number of common factors to be specified manually.

\begin{lstlisting}[language=R]
# Suppose the files are in the same directory as the main script
source("FBPFA-PFA.R")
source("FBPFA-PFA with fixed latent dim.R")
\end{lstlisting}

\textbf{MOM-SS}
The MOM-SS model is implemented in the \texttt{BFR.BE} package, which is available on GitHub. The package depends on \texttt{sparseMatrixStats} and the \texttt{mombf} package, both of which must be installed prior to installation. The following code installs the required dependencies and loads the package:

\begin{lstlisting}[language=R]
BiocManager::install("sparseMatrixStats") # Dependency
install.packages("mombf")

install.packages("devtools") 

devtools::install_github("AleAviP/BFR.BE")
library(BFR.BE)
\end{lstlisting}

\textbf{SUFA}

The SUFA package can be installed from GitHub. On Linux systems, installation requires additional system dependencies, such as \texttt{PROJ}, \texttt{sqlite3}, and \texttt{GDAL}, to be available in the system \texttt{PATH}. On Windows, additional updates, particularly for the \texttt{terra} package, may be necessary. During installation, building the package vignettes can be skipped to reduce installation time, as they involve computationally intensive datasets.

\begin{lstlisting}[language=R]

devtools::install_github("noirritchandra/SUFA", build_vignettes = FALSE)
library(SUFA)
\end{lstlisting}

\textbf{Tetris}

As with PFA, the Tetris model is not available as a standalone R package. The required R scripts must be downloaded manually from the GitHub repository (\href{https://github.com/igrabski/tetris/tree/main}{https://github.com/igrabski/tetris/tree/main}) and placed in the same directory as the main analysis script. 

\begin{lstlisting}[language=R]
# Suppose the files are in the same directory as the main script
source("Tetris.R")
\end{lstlisting}

\textbf{Other utility packages}

The following R packages are also used throughout the tutorial for data manipulation, visualization, and matrix operations:

\begin{lstlisting}[language=R]
library(tidyverse) # for data manipulation and visuallization
library(Matrix) #for the bdiag function
\end{lstlisting}

\subsection{Case 1: Nutrition Data Analysis}

\subsubsection{The data}

For the first case study, we analyze data from the Hispanic Community Health Study / Study of Latinos (HCHS / SOL), a large, multi-site cohort study investigating health and dietary habits among Hispanic/Latino adults in the United States. The data comprises 24-hour dietary recall data collected between 2008 and 2011 from 16,415 participants across four sites (Bronx, Chicago, Miami, and San Diego), and six ethnic backgrounds (Mexican, Puerto Rican, Cuban, Dominican, Central American, and South American) \cite{lavange2010sample}.

We focus our analysis on 42 key nutrients selected to 
 best represent the overall diet habits, with an emphasis on those related to cardiovascular health. Following the approach of 
 \cite{de2022shared}, we perform the multi-study techniques by treating individuals from different ethnic backgrounds as separate studies ($S=6$).

This case study has two primary objectives:
i) to estimate nutritional patterns that are both shared and specific across the six ethnic groups, and
ii) to assess the out-of-sample predictive performance of the various integrative factor analysis methods.

The processed dataset is structured as a list of length $S=6$, where each element is a data frame of dimension $N_s \times P$, with $P = 42$ nutrients and
$N_s = (1364,\ 1517,\ 2210,\ 5184,\ 2478,\ 959)$ representing the number of individuals in each ethnic group.



\subsubsection{Pre-processing}

We begin by excluding individuals with missing nutrient intake values, missing background information, or extreme total energy intake values, resulting in a final sample size of $N=10,460$. Subjects reporting negative nutrient intakes are also removed. This results in a final sample of $N=10,460$ individuals.

Nutrient intake values are transformed using a log transformation, $\log(\text{value} + 0.01)$, to improve adherence to normality assumptions.

For each study, the nutrient data are mean-centered column-wise. Note that this step is handled internally by models such as StackFA, IndFA, BMSFA, and Tetris. For MOM-SS, mean-centering is unnecessary because the model explicitly estimates random intercepts.

\begin{lstlisting}[language=R]
Y_list_scaled <- lapply(
  Y_list, function(x) scale(x, center = TRUE, scale = FALSE)
)
Y_mat_scaled <- Y_list_scaled %>% do.call(rbind, .) %>% as.matrix()
\end{lstlisting}

\subsubsection{Model fitting}\label{sec:fit_nutrition}

We fit each method using $K=6$ common factors and $J_s=2$ ethnic background specific factors, which are reasonable upper bounds based on previous analysis\cite{de2022shared}. The following code demonstrates how each model is estimated using default parameters discussed in Section 2:

\begin{lstlisting}[language=R]
# Stack FA
Y_mat =  Y_list %>% do.call(rbind, .) %>% as.matrix()
fit_stackFA <- MSFA::sp_fa(Y_mat, k = 6, scaling = FALSE, centering = TRUE, 
                              control = list(nrun = 10000, burn = 8000))
# Ind FA
fit_indFA <-
      lapply(1:6, function(s){
        j_s = c(8, 8, 8, 8, 8, 8)
        MSFA::sp_fa(Y_list[[s]], k = j_s[s], scaling = FALSE, centering = TRUE,
                    control = list(nrun = 10000, burn = 8000))
      })

# PFA
N_s <- sapply(Y_list, nrow)
fit_PFA <- PFA(Y=t(Y_mat_scaled),
                        latentdim = 6,
                        grpind = rep(1:6,
                                    times = N_s),
                Cutoff = 0.001,
                Thin = 5,
                Total_itr = 10000, burn = 8000)

# MOM-SS
Y_mat =  Y_list %>% do.call(rbind, .) %>% as.matrix()
# Construct the membership matrix
N_s <- sapply(Y_list, nrow)
M_list <- list()
   for(s in 1:6){
     M_list[[s]] <- matrix(1, nrow = N_s[s], ncol = 1)
   }
M <- as.matrix(bdiag(M_list))
fit_MOMSS <- BFR.BE::BFR.BE.EM.CV(x = Y_mat, v = NULL, 
                                  b = M, q = 6, scaling = FALSE)


# SUFA
fit_SUFA <- SUFA::fit_SUFA(Y_list_scaled, qmax=6, nrun = 10000)

# BMSFA
fit_BMSFA <- MSFA::sp_msfa(Y_list, k = 6, j_s = c(2, 2, 2, 2, 2, 2),
                           outputlevel = 1, scaling = FALSE, 
                           centering = TRUE,
                           control = list(nrun = 10000, burn = 8000))
\end{lstlisting}


Unlike the other methods, fitting Tetris involves a three-step procedure. Posterior samples of factor loadings are not directly comparable between different realizations of the factor sharing matrix $\mathcal{T}$, which complicates direct inference. As recommended by the authors, we first use the tetris function to sample from the posterior distribution of all model parameters, including $\mathcal{T}$. We then apply the \texttt{choose.A} function to obtain a point estimate of $\mathcal{T}$. Finally, we rerun the sampler with $\mathcal{T}$ fixed to this estimate to generate posterior samples for factor loadings.

\begin{lstlisting}[language=R]
# Tetris
set_alpha <- ceiling(1.25*6)
fit_Tetris <- tetris(Y_list, alpha=set_alpha, beta=1, nprint = 200, 
                     nrun=10000, burn=8000)
big_T <- choose.A(fit_Tetris, alpha_IBP=set_alpha, S=6)
run_fixed <- tetris(Y_list, alpha=set_alpha, beta=1, 
                    fixed=TRUE, A_fixed=big_T, nprint = 200, 
                    nrun=10000, burn=8000)
\end{lstlisting}

In practice, model fitting should be conducted in a high-performance computing environment, as some methods are computationally demanding. In our experiments, PFA required more than 10 hours to complete, while Tetris took approximately 4 days to run on a high-performance computing cluster at Brown University. This cluster, managed by the Center for Computation and Visualization, consists of 388 compute nodes with a total of 20,176 CPU cores. In contrast, all other models were completed in under 30 minutes. 

\subsubsection{Post-processing}\label{sec:post-nutrition}

This section describes the post-processing steps used to estimate the number of factors and obtain point estimates for the factor loadings for each method.

\paragraph{Stack FA and Ind FA}
For both Stack FA and Ind FA, post-processing begins by extracting the posterior samples of the loading matrices. To obtain a representative point estimate of the loading matrix $\Phi$, we apply orthogonal Procrustes (OP) alignment across the posterior samples. The common covariance matrix is then estimated as $\Phi\Phi^\top$, while the marginal covariance matrix is computed as the average across posterior draws.

\begin{lstlisting}[language=R]
post_stackFA <- function(fit, S){
  est_Phi <- MSFA::sp_OP(fit$Lambda, trace=FALSE)$Phi
  est_SigmaPhi <- tcrossprod(est_Phi)
  est_SigmaMarginal <-  lapply(1:S, function(s)
    apply(fit$Sigma, c(1, 2), mean)
  )
  Psi_chain <- list()
  for(i in 1:dim(fit$Sigma)[3]){
    Psi_chain[[i]] <- fit$Sigma[, , i] - tcrossprod(fit$Lambda[, , i])
  }
  est_Psi <- Reduce('+', Psi_chain)/length(Psi_chain)
  return(list(Phi = est_Phi, SigmaPhi = est_SigmaPhi, Psi = est_Psi,
              SigmaMarginal = est_SigmaMarginal))
}
res_stackFA <- post_stackFA(fit_stackFA, S=6)
saveRDS(res_stackFA, "Data/Rnutrition_StackFA.rds")
\end{lstlisting}

To estimate the number of latent factors, we perform eigenvalue decomposition (EVD) on the marginal covariance matrix and retain the number of components required to explain a pre-specified proportion of variance. Based on this estimate, the models are refit using the selected number of factors, and the final results are then obtained.

\begin{lstlisting}[language=R]
fun_eigen <- function(Sig_mean) {
  val_eigen <- eigen(Sig_mean)$values
  prop_var <- val_eigen/sum(val_eigen)
  choose_K <- length(which(prop_var > 0.05))
  return(choose_K)
}
res_stackFA <- readRDS("Data/Rnutrition_StackFA.rds")
SigmaPhi_StackFA <- res_stackFA$SigmaPhi 
K_StackFA <- fun_eigen(SigmaPhi_StackFA)
\end{lstlisting}

The estimated number of common factors for Stack FA is $K=4$.

The model is then re-fit using this value, and the final posterior summaries are extracted:

\begin{lstlisting}[language=R]
fit_stackFA_2 <- MSFA::sp_fa(Y_mat_scaled, k = K_StackFA, scaling = FALSE, centering = TRUE, 
                           control = list(nrun = 10000, burn = 8000))
res_stackFA_2 <- post_stackFA(fit_stackFA_2, S=6)
saveRDS(res_stackFA_2, "Data/Rnutrition_StackFA_2.rds")
\end{lstlisting}

A similar procedure is applied for Ind FA. First, we post-process the output to estimate the factor loadings, marginal covariance matrices, and noise variances:

\begin{lstlisting}[language=R]
# Ind FA
post_indFA <- function(fit){
  # Estimated study-specific covariance and loading
  S <- length(fit_list)
  est_LambdaList <- lapply(1:S, function(s){
    MSFA::sp_OP(fit_list[[s]]$Lambda, trace=FALSE)$Phi
  })
  est_SigmaLambdaList <- lapply(est_LambdaList, function(x) tcrossprod(x))
  
  # Marginal covariance matrices
  est_SigmaMarginal <- lapply(1:S, function(s) {
    fit <- fit_list[[s]]
    apply(fit$Sigma, c(1, 2), mean)
  })

  Psi <- list()
  for(s in 1:S){
    Psi_chain <- list()
    for(i in 1:dim(fit_list[[1]]$Sigma)[3]){
      Psi_chain[[i]] <- fit_list[[s]]$Sigma[, , i] - tcrossprod(fit_list[[s]]$Lambda[, , i])
    }
    Psi[[s]] <- Reduce('+', Psi_chain)/length(Psi_chain)
  }
  
  return(list(LambdaList = est_LambdaList, SigmaLambdaList = est_SigmaLambdaList, Psi = Psi,
              SigmaMarginal = est_SigmaMarginal))
}

res_indFA <- post_indFA(fit_indFA)
saveRDS(res_indFA, "Data/Rnutrition_IndFA.rds")
\end{lstlisting}

The number of factors for each study is estimated by applying eigenvalue decomposition to the covariance matrices:

\begin{lstlisting}[language=R]
SigmaLambda_IndFA <- readRDS("Data/Rnutrition_IndFA.rds")$SigmaLambda
Js_IndFA <- lapply(SigmaLambda_IndFA, fun_eigen)
\end{lstlisting}

The estimated values of $J_s$ for Ind FA are 4, 5, 5, 5, 4, and 4. 

The model is then re-fit for each study using the corresponding number of factors:

\begin{lstlisting}[language=R]
# Ind FA
fit_indFA_2 <-
  lapply(1:6, function(s){
    j_s = c(4, 5, 5, 5, 4, 4)
    MSFA::sp_fa(Y_list[[s]], k = j_s[s], scaling = FALSE, centering = TRUE,
                control = list(nrun = 10000, burn = 8000))
  })
res_indFA_2 <- post_indFA(fit_indFA_2)
saveRDS(res_indFA_2, "Data/Rnutrition_IndFA_2.rds")
\end{lstlisting}

\paragraph{PFA}
The following code performs post-processing for the PFA model.

We first estimate the number of factors $K$ by calculating the number of columns in the loading matrix from each MCMC sample. We identify the mode of these values and retain only those posterior samples corresponding to this modal 
$K$ for downstream analysis.

To obtain the common loading matrix, we multiply the loading matrix  $\Phi$ by the square root of the latent variance matrix $V^{1/2}$  for each retained sample.  Although the average of these transformed matrices $\Phi V^{1/2}$can be used as a point estimate, we apply orthogonal Procrustes (OP) rotation to account for identifiability issues. For the common covariance, we compute $\Phi V\Phi^\top$ at each iterations, and verage across samples. Similar procedure is applied for the study-specific and marginal covariance matrices.

\begin{lstlisting}[language=R]
post_PFA <- function(fit) {
  # Determine posterior dimension (number of factors per sample)
  k_vec <- sapply(fit$Loading, ncol)
  mode_k <- as.numeric(names(sort(table(k_vec), decreasing = TRUE)[1]))
  
  # Filter posterior samples to those with mode_k
  keep_idx <- which(k_vec == mode_k)
  fit$Loading <- fit$Loading[keep_idx]
  fit$Latentsigma <- fit$Latentsigma[keep_idx]
  fit$Errorsigma <- fit$Errorsigma[keep_idx]
  fit$Pertmat <- fit$Pertmat[keep_idx]
  
  npost <- length(fit$Loading)
  p <- nrow(fit$Loading[[1]])
  k <- mode_k
  S <- dim(fit$Pertmat[[1]])[2]
  
  posteriorPhis <- array(0, dim = c(p, k, npost))
  posteriorLams <- vector("list", S)

  for(s in 1:S){
    posteriorLams[[s]] <- array(0, dim = c(p, k, npost))
    for(i in 1:npost){
      posteriorPhis[,,i] <- fit$Loading[[i]] %*% diag(fit$Latentsigma[[i]])
      posteriorLams[[s]][,,i] <- (solve(matrix(fit$Pertmat[[i]][, s], p, p)) - diag(p)) %*% posteriorPhis[,,i]
    }
  }

  # Varimax rotation
  est_Phi <- MSFA::sp_OP(posteriorPhis, itermax = 10, trace = FALSE)$Phi
  est_speLoad <- lapply(posteriorLams, function(x) MSFA::sp_OP(x, itermax = 10, trace = FALSE)$Phi)

  # Estimated covariance components
  sharevar <- list()
  est_SigmaLambdaList <- vector("list", S)
  est_SigmaMarginal <- vector("list", S)
  est_Psi_list <- list()

  for(s in 1:S){
    post_SigmaLambda_s <- vector("list", npost)
    post_SigmaMarginal_s <- vector("list", npost)
    Psi <- vector("list", npost)

    for(i in 1:npost){
      sharevar[[i]] <- fit$Loading[[i]] %*% diag(fit$Latentsigma[[i]]^2) %*% t(fit$Loading[[i]]) + 
        diag(fit$Errorsigma[[i]]^2)
      Q_temp_inv <- solve(matrix(fit$Pertmat[[i]][, s], p, p))
      post_SigmaMarginal_s[[i]] <- Q_temp_inv %*% sharevar[[i]] %*% t(Q_temp_inv)
      post_SigmaLambda_s[[i]] <- post_SigmaMarginal_s[[i]] - sharevar[[i]]
      Psi[[i]] <- diag(fit$Errorsigma[[i]]^2)
    }

    est_SigmaMarginal[[s]] <- Reduce('+', post_SigmaMarginal_s) / npost
    est_SigmaLambdaList[[s]] <- Reduce('+', post_SigmaLambda_s) / npost
    est_Psi_list[[s]] <- Reduce('+', Psi) / npost
  }

  est_Psi <- Reduce('+', est_Psi_list) / S
  est_SigmaPhi <- Reduce('+', sharevar) / npost
  est_Q <- Reduce('+', fit$Pertmat) / npost
  est_Q_list <- lapply(1:S, function(s) matrix(est_Q[, s], p, p))

  return(list(
    Phi = est_Phi,
    SigmaPhi = est_SigmaPhi,
    Psi = est_Psi,
    Q = est_Q_list,
    LambdaList = est_speLoad,
    SigmaLambdaList = est_SigmaLambdaList,
    SigmaMarginal = est_SigmaMarginal,
    mode_k = mode_k,
    kept_samples = length(keep_idx)
  ))
}
res_PFA <- post_PFA(fit_PFA)
saveRDS(res_PFA, "Data/Rnutrition_PFA.rds") 
\end{lstlisting}

\paragraph{MOM-SS}
For MOM-SS, the common factor loading matrix  $\Phi$ is directly obtained from the fitted output as a post-processed estimate. The common covariance is computed as $\Phi\Phi^\top$. The marginal covariance matrix $\Sigma_{s}$ is then calculated by adding the estimated study-specific residual covariance matrices to the common covariance component. Additionally, the study-specific intercepts $\alpha$ and the regression coefficients for the known covariates $B$ are  extracted from the fitted object.

\begin{lstlisting}[language=R]
post_MOMSS <- function(fit, version = 2){ # version 1: M, version 2: Mpost
  est_Phi <- fit$M
  if (version==2){est_Phi <- fit$Mpost}
  est_SigmaPhi <- tcrossprod(est_Phi)
  
  # Marginal covariance
  S <- dim(fit$sigma)[2]
  est_PsiList <- est_SigmaMarginal <-  list()
  for(s in 1:S){
    est_PsiList[[s]] <- fit$sigma[,s]
    est_SigmaMarginal[[s]] <- est_SigmaPhi + diag(fit$sigma[,s])
  }
  # last S columns of fit$Theta are the study-specific intercepts
  est_alphas <- fit$Theta[, (dim(fit$Theta)[2]-S+1):dim(fit$Theta)[2]]
  # The rest are coeficients for the known covariates
  est_B <- fit$Theta[, 1:(dim(fit$Theta)[2]-S)]
  
  return(list(Phi = est_Phi, SigmaPhi = est_SigmaPhi, Psi = est_PsiList, alpha = est_alphas, B = est_B,
              SigmaMarginal = est_SigmaMarginal))
}
res_MOMSS <- post_MOMSS(fit_MOMSS)
saveRDS(res_MOMSS, "Data/Rnutrition_MOMSS.rds")
\end{lstlisting}

\paragraph{SUFA}
For SUFA, the common and study-specific factor loading matrices, along with the common and marginal covariance matrices are obtained using the \texttt{lam.est.all()}, \texttt{SUFA\_shared\_covmat()} and \texttt{sufa\_marginal\_covs()} functions. The residual covariance is computed by averaging the residuals extracted from the function's output. Study-specific covariance matrices are calculated by subtracting the common covariance from the marginal covariance. It is important to note that, under the SUFA model formulation, the common covariance matrix is defined as  $\Phi\Phi^\top + \Sigma$.

\begin{lstlisting}[language=R]
post_SUFA <- function(fit){
  all <- dim(fit$Lambda)[3]
  burnin <- floor(all * 0.8) # We will use the last 20% samples
  # shared and study-specific loading matrices
  loadings <- lam.est.all(fit, burn = burnin)
  # Obtain common covariance matrix and loading from fitting
  est_Phi <- loadings$Shared
  est_SigmaPhi <- SUFA_shared_covmat(fit, burn = burnin)
  est_Psi <- diag(colMeans(fit$residuals))
  # Study-specific loadings
  est_LambdaList <- loadings$Study_specific
  
  # Obtain study-specific covariance matrices
  S <- length(fit$A)
  marginal_cov <- sufa_marginal_covs(fit, burn = burnin)
  est_SigmaLambdaList <- list()
  for (s in 1:S) {
    est_SigmaLambdaList[[s]] <- marginal_cov[,,s] - est_SigmaPhi
  }
  
  return(list(SigmaPhi = est_SigmaPhi, Phi = est_Phi, 
              SigmaLambdaList = est_SigmaLambdaList,
              LambdaList = est_LambdaList, 
              Psi = est_Psi,
              SigmaMarginal = lapply(1:S, function(s) marginal_cov[,,s])
              ))
}
res_SUFA <- post_SUFA(fit_SUFA)
saveRDS(res_SUFA, "Data/Rnutrition_SUFA.rds")
\end{lstlisting}

\paragraph{BMSFA}
For BMSFA, the post-processing procedure follows the same steps as described for Stack FA and Ind FA:

\begin{lstlisting}[language=R]
post_BMSFA <- function(fit){
  # Common covariance matrix and loading
  est_Phi <- sp_OP(fit$Phi, trace=FALSE)$Phi
  est_SigmaPhi <- tcrossprod(est_Phi)
  
  # Study-specific covariance matrices and loadings
  est_LambdaList <- lapply(fit$Lambda, function(x) sp_OP(x, trace=FALSE)$Phi)
  est_SigmaLambdaList <- lapply(est_LambdaList, function(x) tcrossprod(x))
  
  # Marginal covariance matrices
  S <- length(est_SigmaLambdaList)
  # Get point estimate of each Psi_s
  est_PsiList <- lapply(1:S, function(s) {
    apply(fit$psi[[s]], c(1, 2), mean)
  })
  est_margin_cov <- lapply(1:S, function(s) {
    est_SigmaPhi + est_SigmaLambdaList[[s]] + diag(est_PsiList[[s]] %>% as.vector())
  })
  
  return(list(Phi = est_Phi, SigmaPhi = est_SigmaPhi,
         LambdaList = est_LambdaList, SigmaLambdaList = est_SigmaLambdaList,
         PsiList = est_PsiList,
         SigmaMarginal = est_margin_cov))
}
res_BMSFA <- post_BMSFA(fit_BMSFA)
saveRDS(res_BMSFA, "Data/Rnutrition_BMSFA.rds")
\end{lstlisting}

We estimate the number of factors by applying eigenvalue decomposition to the common and study-specific covariance matrices:
\begin{lstlisting}[language=R]
SigmaPhi_BMSFA <- readRDS("Data/Rnutrition_BMSFA.rds")$SigmaPhi
K_BMSFA <- fun_eigen(SigmaPhi_BMSFA)
SigmaLambda_BMSFA <- readRDS("Data/Rnutrition_BMSFA.rds")$SigmaLambda
Js_BMSFA <- lapply(SigmaLambda_BMSFA, fun_eigen)
\end{lstlisting}

We then re-run the model using these values:
The estimated number of common factors is  $K=4$,  and the study-specific numbers of factors are $J_s = 2, s=1, \cdots, 6$. We then re-run the model using these values:
\begin{lstlisting}[language=R]
fit_BMSFA_2 <- MSFA::sp_msfa(Y_list, k = 4, j_s = c(2, 2, 2, 2, 2, 2),
                           outputlevel = 1, scaling = FALSE, 
                           centering = TRUE,
                           control = list(nrun = 10000, burn = 8000))
res_BMSFA_2 <- post_BMSFA(fit_BMSFA_2)
saveRDS(res_BMSFA_2, "Data/Rnutrition_BMSFA_2.rds")
\end{lstlisting}

\paragraph{Tetris}
For Tetris, the combinatorial loadings $\Phi^{*}$ are obtained using the \texttt{getLambda()} function. The common factor loadings $\Phi$ are computed as $\Phi^{*}P$, and the study-specific factor loadings $\Lambda_s$ as $\Phi^{*}R_s$, where $P + R_s =T_s$, following the definition in Table \ref{tab:summary}. The common covariance matrix is computed as $\Phi\Phi^\top$ and the study-specific covariance matrices as $\Lambda_s\Lambda_s^\top$, and the marginal covariance matrices as $\Phi T_s \Phi^\top + \Psi_s$, for each $s=1,\cdots, S$.

\begin{lstlisting}[language=R]
 post_Tetris <- function(fit){
  # Estimated common covariance
  A <- fit$A[[1]]
  Lambda <- getLambda(fit,A)
  S <- dim(A)[1]
  est_Phi <- as.matrix(Lambda[,colSums(A)==S])
  est_SigmaPhi <- tcrossprod(est_Phi)
  # Estimated study-specific covariance
  P = diag((colSums(A) == S)*1)
  T_s <- list()
  est_LambdaList <- list()
  for(s in 1:S){
    T_s[[s]] <- diag(A[s,])
    Lambda_s <- Lambda %*% (T_s[[s]] - P)
    Lambda_s <- Lambda_s[,-which(colSums(Lambda_s == 0) == nrow(Lambda_s))]
    Lambda_s <- matrix(Lambda_s, nrow=nrow(Lambda))
    est_LambdaList[[s]] <- Lambda_s}
  est_SigmaLambdaList <- lapply(1:S, function(s){
    tcrossprod(est_LambdaList[[s]])})
  
  # Estimated marginal covariance
  Psi <- list()
  est_SigmaMarginal <- lapply(1:S, function(s){
    Psi[[s]] <- diag(Reduce("+", fit$Psi[[s]])/length(fit$Psi[[s]]))
    Sigma_s <- Lambda %*% T_s[[s]] %*% t(Lambda) + Psi[[s]]
    })
  
  return(list(Phi = est_Phi, SigmaPhi = est_SigmaPhi,
              LambdaList = est_LambdaList, SigmaLambdaList = est_SigmaLambdaList,
              Psi = Psi, T_s = T_s,
              SigmaMarginal = est_SigmaMarginal))
}
res_Tetris <- post_Tetris(run_fixed)
saveRDS(res_Tetris, "Data/Rnutrition_Tetris.rds")
\end{lstlisting}



\subsubsection{Visualization}

We visualize the estimated factor loadings using heatmaps, where each row represents a variable (i.e., nutrient) and each column corresponds to a latent factor. To ensure consistency across methods, columns in each loading matrix are ordered according to the  proportion of variance explained by each factor, computed as the corresponding eigenvalue divided by the sum of the eigenvalue.  

\begin{figure}[!ht]
\begin{subfigure}{.28\textwidth}
  \centering
  \includegraphics[width=.95\linewidth]{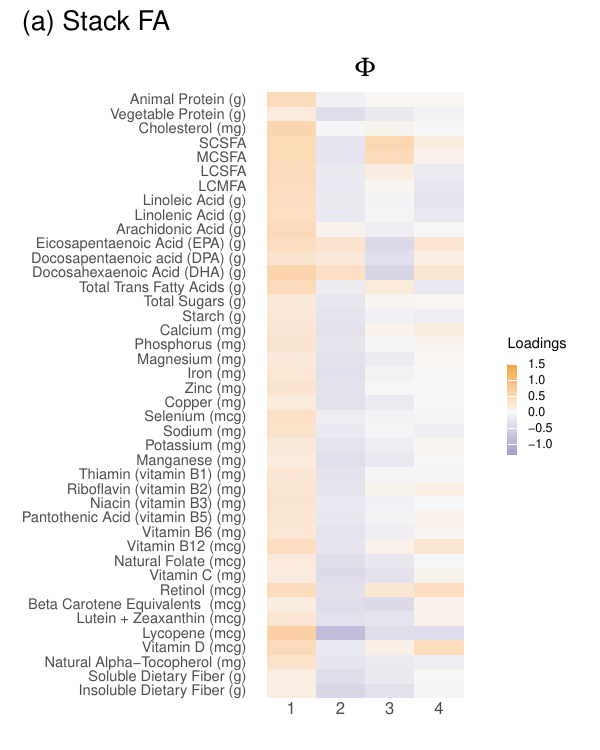}
  \label{fig:sfiga}
\end{subfigure}%
\begin{subfigure}{.71\textwidth}
  \centering
  \includegraphics[width=.95\linewidth]{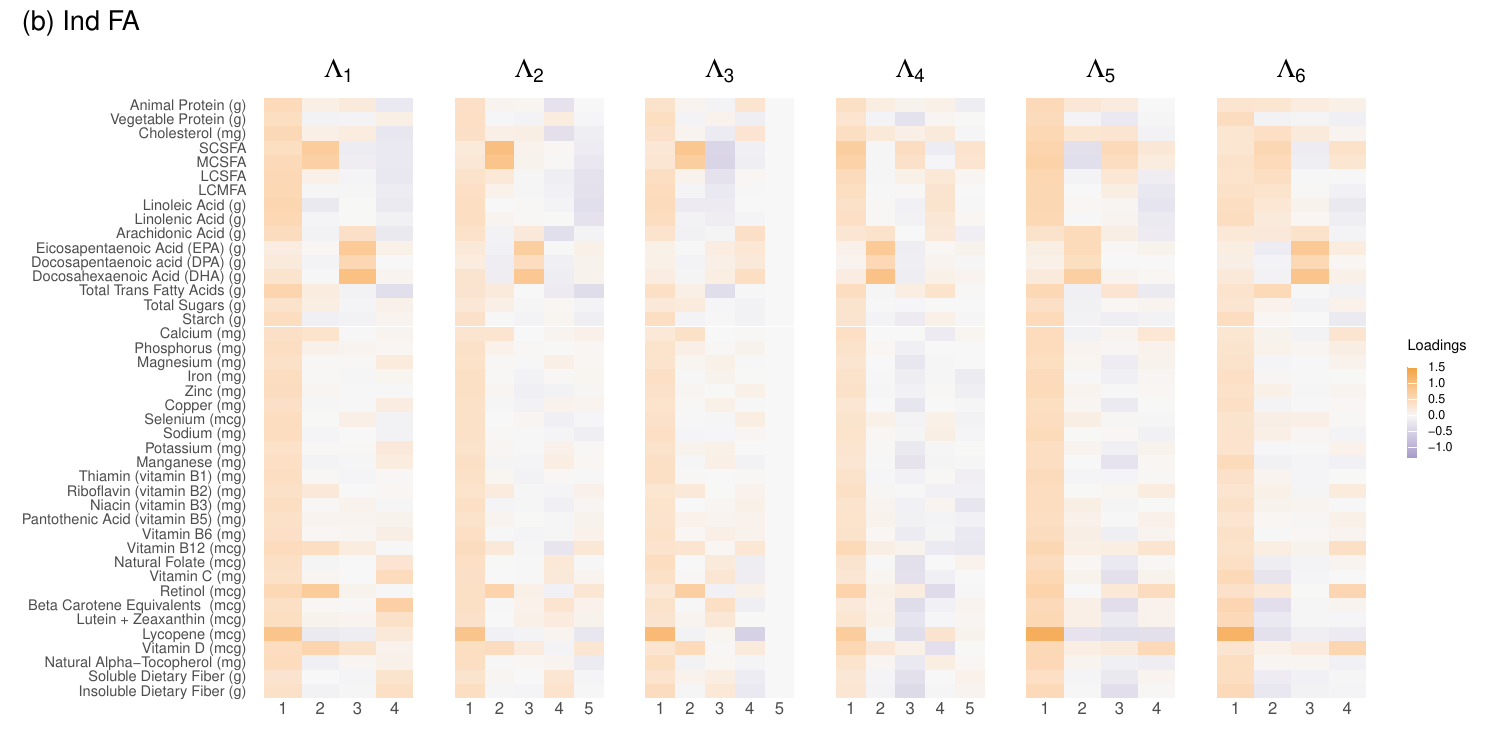}
  \label{fig:sfigb}
\end{subfigure}
\begin{subfigure}{.27\textwidth}
    \centering
    \includegraphics[width=.95\linewidth]{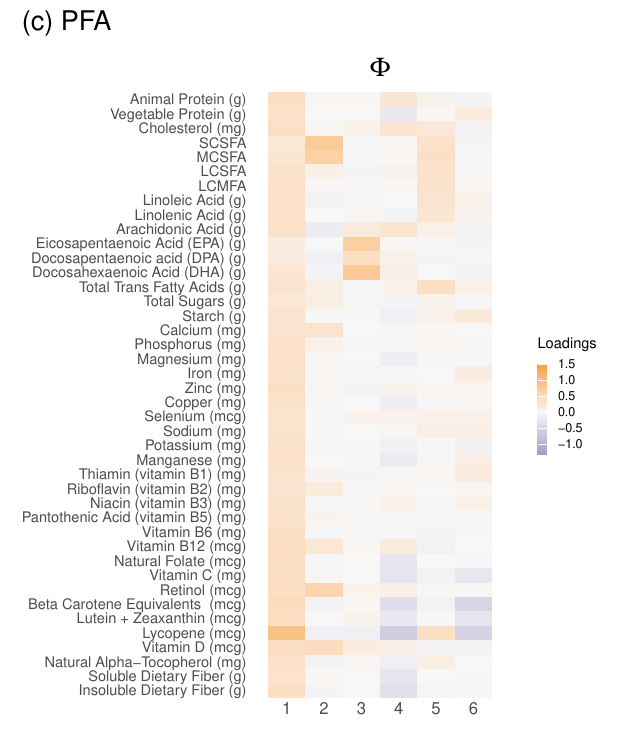}
    \label{fig:sfigc}
\end{subfigure}
\begin{subfigure}{.27\textwidth}
    \centering
    \includegraphics[width=.95\linewidth]{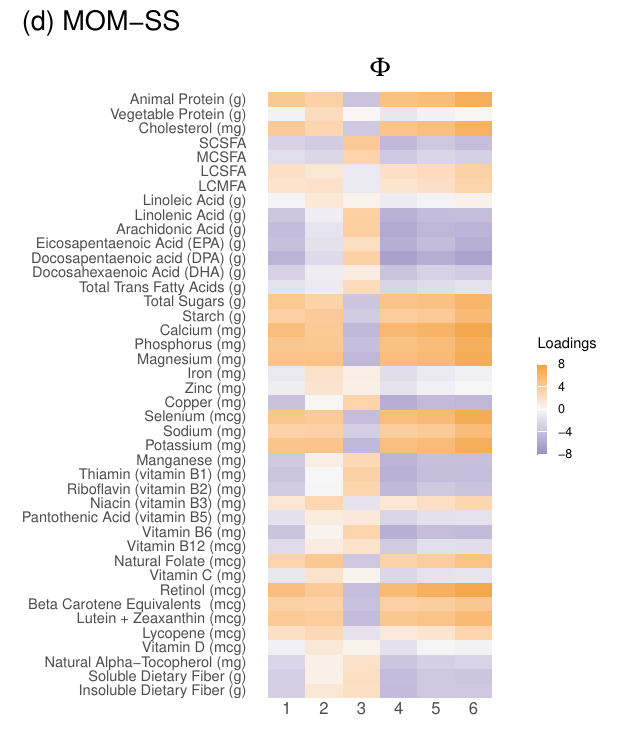}
    \label{fig:sfigd}
\end{subfigure}
\begin{subfigure}{.44\textwidth}
    \centering
    \includegraphics[width=.95\linewidth]{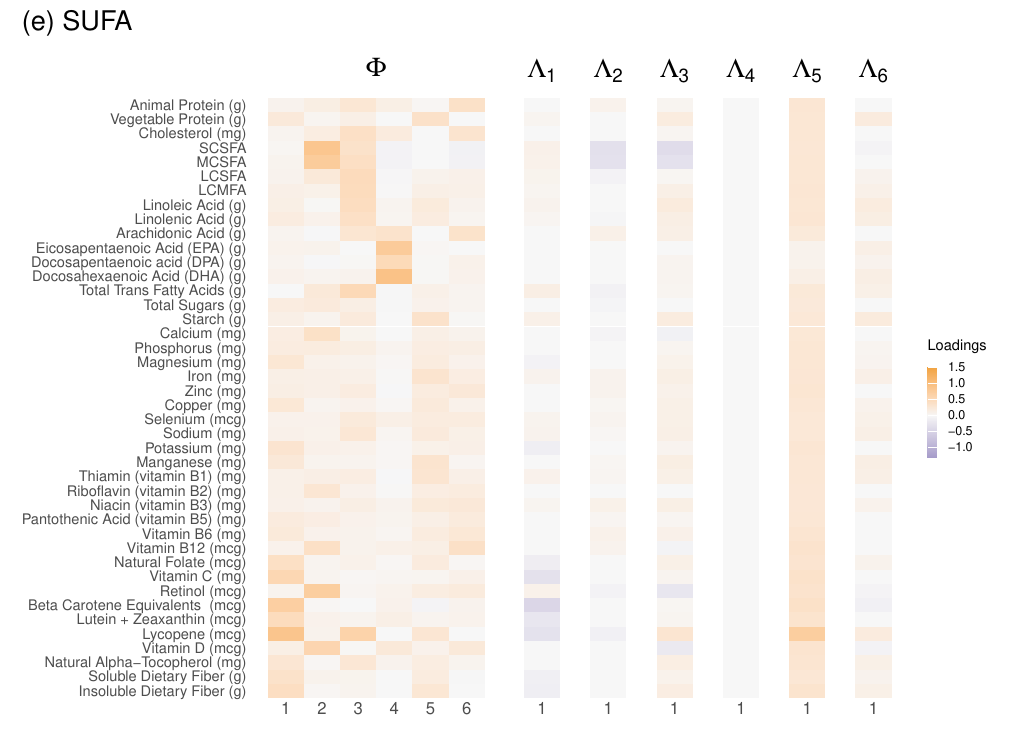}
    \label{fig:sfige}
\end{subfigure}\\
\begin{subfigure}{.49\textwidth}
    \centering
    \includegraphics[width=.95\linewidth]{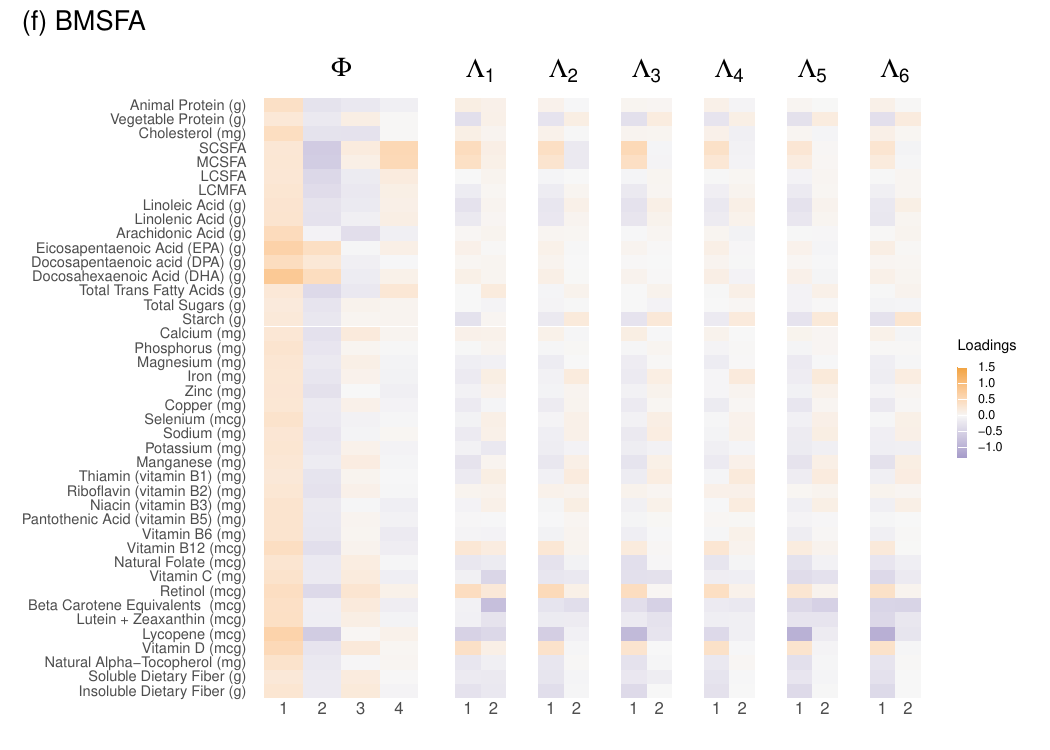}
    \label{fig:sfigf}
\end{subfigure}
\begin{subfigure}{.5\textwidth}
    \centering
    \includegraphics[width=.95\linewidth]{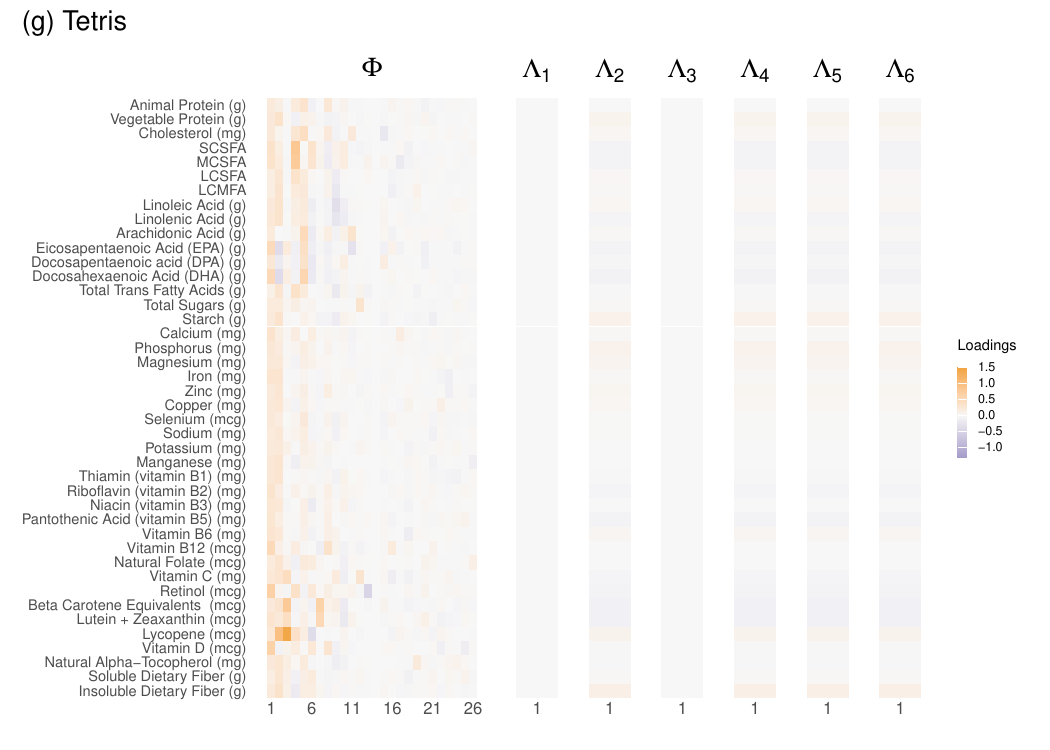}
    \label{fig:sfigg}
\end{subfigure}
\caption{\it Heatmaps of the factor loading matrices for each method. Rows correspond to nutrients, and columns  to  latent factors. Loadings in blue (orange) represent negative (positive) associations.}
\label{fig:heat}
\end{figure}

The estimated factor loadings across methods reveal consistent dietary patterns. Factors with high positive loadings on animal protein, saturated fat, and cholesterol likely reflect meat-heavy dietary habits, whereas those with strong positive loadings on fiber, folate, and plant-based proteins suggest a plant-based diets. Similarly, factors dominated by omega-3 fatty acid and nutrients associated with seafood consumption indicate seafood-based dietary patterns.

The methods differ notably in the level of sparsity and thus interpretability.  Stack FA, Ind FA, and PFA produce clear loading structures, where nutrient groupings are well defined and associated with distinct dietary profiles. In contrast, MOM-SS estimates more diffuse loadings, with multiple nutrients contributing to each factor, making interpretation more complex. This diffuse structure aligns with the behavior of MOM-SS’s non-local spike-and-slab priors, which encourage a sharp separation between near-zero and non-zero loadings. SUFA and BMSFA impose greater sparsity, resulting in more interpretable loadings in which each factor is primarily driven by a small number of nutrients. Tetris, on the other hand, tends  to overestimate the number of factors, producing in nearly empty  loaded columns.

The Tetris method required extensive computational time, taking over four days to complete. This is due not only to the large sample size ($N$) but also to its iterative search for  additional common factors. Similar behavior was observed in Simulation Scenario 1, where Tetris overestimated the number of factors under a data-generating process consistent with a PFA model.
To improve computational efficiency in such settings, one may specify a fixed number of factors (i.e., a fixed $\mathcal{T}$) or adjust the inclusion parameters ($\alpha_\mathcal{T}$ and $\beta_\mathcal{T}$). 
 In scenarios such as the nutrition data or Simulation Scenario 1, where the number of observations ($N$) exceeds the number of variables ($P$), smaller values of  $\alpha_\mathcal{T}$ and $\beta_\mathcal{T}$, are recommended to prevent excessive number of factor.

\subsubsection{Model Prediction Accuracy}

To evaluate the predictive performance  of each model, we compute the mean squared error (MSE)  by reconstructing observed data from  estimated factor scores and loadings. Each model is trained on a randomly selected 70\% of the data, and prediction accuracy is evaluated on the remaining 30\%. Predictions are obtained as:
\begin{equation}
\label{eq:MSE-stackFA}
\begin{array}{cl}
     \widehat{\mathbf{f}}_{is,(new)}&=(\widehat{\Phi}^\top\widehat{\Psi}^{-1}\widehat{\Phi})^{-1}\widehat{\Phi}^\top\widehat{\Psi}^{-1}\mathbf{y}_{is,(new)}\,,\\  \widehat{\mathbf{y}}_{is,(new)} &= \widehat{\Phi} \widehat{\mathbf{f}}_{is,(new)},
\end{array}
\end{equation}
where $\widehat{\Phi}$ denotes the factor loadings matrix estimated from the training data (using Stack FA in this illustrative example), and $\widehat{\mathbf{f}}_{is,(new)}$ is the corresponding factor score 
estimated using an adaptation of the Bartlett method \cite{bartlett1937statistical, hansen2024fast}. The mean squared error of prediction is then calculated as:
\begin{equation}
    \frac{1}{P\sum _s^S N_s}\sum_{s}^S\sum_{i}^{N_s}\sum_{p}^P(\widehat{y}_{isp,(new)}-y_{isp,(new)})^2,
\end{equation}

using the 30\% of samples in each study that were held out for testing.

\begin{lstlisting}[language=R]
train_ratio <- 0.7
train_list <- list()
test_list <- list()

set.seed(6)
for (s in seq_along(Y_list)) {
  N_s <- nrow(Y_list[[s]])  # Number of rows in the study
  train_indices <- sample(1:N_s, size = floor(train_ratio * N_s), replace = FALSE)
  
  train_list[[s]] <- Y_list[[s]][train_indices, ]
  test_list[[s]] <- Y_list[[s]][-train_indices, ]
}
# Test data has to be centered
test_list <- lapply(test_list, as.matrix)
test_list_scaled <- lapply(
  test_list, function(x) scale(x, center = TRUE, scale = FALSE)
)
\end{lstlisting}

Using the same example, the MSE for Stack FA is computed as follows:

\begin{lstlisting}[language=R]
# Stack FA
Phi <- fit_StackFA_train$Phi
Psi <- fit_StackFA_train$Psi

mse_stackFA <- 1/(42 * sum(sapply(test_list, nrow)))*sum(
  sapply(1:6, function(s){
    scores <- test_list_scaled[[s]] %*% solve(Psi) %*% Phi %*% mnormt::pd.solve(signif(t(Phi) %*% solve(Psi) %*% Phi))
    norm(test_list_scaled[[s]] - scores %*% t(Phi), "F")^2
  })
)
\end{lstlisting}

Table~\ref{tab:mse_nutrition} reports the MSE for each method.

\begin{table}[htbp]
\centering
\caption{\it Mean Squared Error (MSE) of different models.}
\renewcommand{\arraystretch}{1.2}
\begin{tabular}{l c c ccccc}
\toprule
\textbf{Model} & Stack FA & Ind FA & PFA & MOM-SS &SUFA & BMSFA & Tetris  \\
\midrule
 \textbf{MSE} & 0.514 &  0.503 & 0.706 & 0.490 & 0.462 & 0.473 & 0.318 \\
\bottomrule
\end{tabular}
\label{tab:mse_nutrition}
\end{table}

The MSE values provide insight into how well each model perform prediction. As expected, Tetris achieves the lowest MSE (0.314), due to its estimation of a large number of factors ($K=26$). However, as observed in the loading matrices, many of these factors presents very small or absent loadings, suggesting that Tetris overfits the data.  Following TETRIS, SUFA (0.462) and BMSFA (0.473) yield the lowest prediction errors among the remaining models,
with a balance between low error and model parsimony. These results highlight their effectiveness in capturing both common and study-specific dietary patterns.

Conversely, PFA shows the highest MSE (0.706), indicating limited predictive accuracy—likely a consequence of its structural constraint requiring all studies to align with a common reference loading, which may fail to accommodate population-specific dietary variability.

\subsection{Case 2: Gene Expression Data Analysis}

In this demonstration, we use the curatedOvarianData package \cite{ganzfried2013curatedovariandata} to illustrate (1)  common gene co-expression network captured by the shared covariance matrix $\Sigma_\Phi$, and (2) model performance through prediction error, measured by mean squared error (MSE).  This dataset includes gene expression microarray data and clinical outcomes for 2,970 ovarian cancer patients across 23 studies. The studies vary in terms of sequencing platform, sample size, tumor stage and subtype, survival, and censoring information.

\subsubsection{Loading the data}

We begin by loading the \texttt{curatedOvarianData} package, which contains standardized and preprocessed expression data from multiple ovarian cancer studies. A list of all available datasets can be retrieved using:

\begin{lstlisting}[language=R]
library(curatedOvarianData)
data(package="curatedOvarianData")    
\end{lstlisting}

For our analysis, we select four representative studies:

\begin{lstlisting}[language=R]
data(GSE13876_eset)
data(GSE26712_eset)
data(GSE9891_eset)
data(PMID17290060_eset)
\end{lstlisting}

These four datasets have a similar sample size. Across all four studies, the majority of patients are in advanced cancer stages, and the predominant histological subtype is serous carcinoma. However, they differ in sequencing platforms: Operon V3 two-color, Affymetrix HG-U133A, Affymetrix HG-U133 Plus 2.0, and Affymetrix HG-U133A, respectively.

\subsubsection{Pre-processing}

First we identify  the genes common in the four studies. 
 The \texttt{featureNames} function is used to extract gene names, and \texttt{exprs} retrieves the expression matrices \cite{falcon2007introduction}.
\begin{lstlisting}[language=R]
inter_genes <- Reduce(intersect, list(featureNames(GSE13876_eset),
                                      featureNames(GSE26712_eset), 
                                      featureNames(GSE9891_eset),
                                      featureNames(PMID17290060_eset)))

GSE13876_eset <- GSE13876_eset[inter_genes,]
GSE26712_eset <- GSE26712_eset[inter_genes,]
GSE9891_eset <- GSE9891_eset[inter_genes,]
PMID17290060_eset <- PMID17290060_eset[inter_genes,]

study1 <- t(exprs(GSE13876_eset))
study2 <- t(exprs(GSE26712_eset))
study3 <- t(exprs(GSE9891_eset))
study4 <- t(exprs(PMID17290060_eset))
\end{lstlisting}

Next, we filter the genes with high variance by using the coefficient of variation (CV). Genes with a CV above a fixed threshold (0.16) in at least one study are retained:

\begin{lstlisting}[language=R]
# calculate Coefficient of Variation of each gene
cv1 <- apply(study1, 2, sd) / apply(study1, 2, mean)
cv2 <- apply(study2, 2, sd) / apply(study2, 2, mean)
cv3 <- apply(study3, 2, sd) / apply(study3, 2, mean)
cv4 <- apply(study4, 2, sd) / apply(study4, 2, mean)
cv_matrix <- rbind(cv1, cv2, cv3, cv4)

# Find genes with CV >= threshold in at least one study
threshold <- 0.16
genes_to_keep <- apply(cv_matrix, 2, function(cv) any(cv >= threshold))
sum(genes_to_keep)
# Filtered
study1 <- GSE13876_eset[genes_to_keep,]
study2 <- GSE26712_eset[genes_to_keep,]
study3 <- GSE9891_eset[genes_to_keep,]
study4 <- PMID17290060_eset[genes_to_keep,]
\end{lstlisting}

The expression data are then log-transformed to improve normality and saved in lists of $N_s \times P$ matrices:

\begin{lstlisting}[language=R]
df1 <- study1 %>% exprs() %>% t() %>% 
  log() %>% as.data.frame()
df2 <- study2 %>% exprs() %>% t() %>% 
  log() %>% as.data.frame()
df3 <- study3 %>% exprs() %>% t() %>% 
  log()  %>% as.data.frame()
df4 <- study4 %>% exprs() %>% t() %>% 
  log()  %>% as.data.frame()
list_gene <- list(df1, df2, df3, df4)
saveRDS(list(df1, df2, df3, df4), "./RDS/CuratedOvarian_processed.rds")
\end{lstlisting}




The resulting dataset includes $N_s = (157, 195, 285, 117)$ for $s = 1, 2, 3, 4$, with $P = 1060$ genes.

We standardize the data to focus on correlations across genes. 

\begin{lstlisting}[language=R]
Y_list <- readRDS("./RDS/CuratedOvarian_processed.rds")
Y_list_scaled <- lapply(
 Y_list, function(x) scale(x, center = TRUE, scale = TRUE)
)
\end{lstlisting}

\subsubsection{Model fitting}

We proceed to fit the models. Below is an example using Stack FA. For all other methods, we follow the same fitting approach used in the nutrition data analysis (see Section~\ref{sec:fit_nutrition}) or refer to the full code available at \href{https://mavis-liang.github.io/Bayesian_integrative_FA_tutorial/}{https://mavis-liang.github.io/Bayesian\_integrative\_FA\_tutorial}.

\begin{lstlisting}[language=R]
Y_mat =  Y_list %>% do.call(rbind, .) %>% as.matrix()
fit_stackFA <- MSFA::sp_fa(Y_mat, k = 20, scaling = TRUE, centering = TRUE, 
                              control = list(nrun = 10000, burn = 8000))
\end{lstlisting}

The number of factors is set to $J_s = 20$ for Ind FA, $K = 20$ for MOM-SS, PFA, and SUFA, and $K = 20$, $J_s = 4$ for BMSFA.

\subsubsection{Post-processing}

Post-processing for obtaining point estimates of the loading and covariance matrices follows the same procedure as in the nutrition case. For Stack FA, Ind FA, and BMSFA, this includes an additional step of estimating the number of factors via eigenvalue decomposition (EVD), followed by re-fitting the model with the estimated number.

For example, with Stack FA:
\begin{lstlisting}[language=R]
res_stackFA = post_stackFA(fit_stackFA, S=4)
saveRDS(res_stackFA, "./RDS/results_curated/RCuratedOvarian_stackFA_scaled.rds")
\end{lstlisting}

To estimate the number of factors:
\begin{lstlisting}[language=R]
SigmaPhi_StackFA <- readRDS("./RDS/results_curated/RCuratedOvarian_stackFA_scaled.rds")$SigmaPhi
K_StackFA <- fun_eigen(SigmaPhi_StackFA)
\end{lstlisting}
We then re-run the model with K\_StackFA:

\begin{lstlisting}[language=R]
fit_stackFA <- MSFA::sp_fa(Y_mat, k = K_StackFA, scaling = TRUE, centering = TRUE, 
                              control = list(nrun = 10000, burn = 8000))
res_stackFA = post_stackFA(fit_stackFA, S=4)
saveRDS(res_stackFA, "./RDS/results_curated/RCuratedOvarian_stackFA2_scaled.rds")
\end{lstlisting}

The estimated number of factors for each model is summarized in Table~\ref{tab:table_genomics}.

\subsubsection{Visualization}

Next, we use Gephi\cite{bastian2009gephi} to visualize the common gene co-expression networks using the estimated common covariances. Edges between two genes are included if the absolute value of their corresponding entry in $\widehat{\Sigma}_{\Phi}$ exceeds a threshold. As the estimated covariances differ in magnitude across methods, we apply method-specific thresholds.
 For instance, the thresholds are 0.55 for PFA, 0.85 for Stack FA, 0.95 for MOM-SS, 0.28 for SUFA, and 0.5 for BMSFA. 

\begin{lstlisting}[language=R]
genenames <- colnames(list_gene[[1]])
SigmaPhiPFA <- readRDS("./RDS/results_curated/RCuratedOvarian_PFA_scaled.rds")$SigmaPhi
colnames(SigmaPhiPFA) <- rownames(SigmaPhiPFA) <- genenames
# We ignore the diagonal values
diag(SigmaPhiPFA) <- NA
above_thresh <- abs(SigmaPhiPFA) > 0.55
keep_genes <- apply(above_thresh, 1, function(x) any(x, na.rm = TRUE))
# Entries smaller than 0.55 are set to 0
SigmaPhiPFA[abs(SigmaPhiPFA) < 0.55] <- 0
SigmaPhiPFA_filtered <- SigmaPhiPFA[keep_genes, keep_genes]
write.csv(SigmaPhiPFA_filtered, "SigmaPhiPFA_filtered.csv")
\end{lstlisting}

\begin{figure}[!ht]
\begin{subfigure}{.55\textwidth}
    \centering
    \caption{PFA}
    \includegraphics[width=.95\linewidth,trim={0 3.5cm 0cm 2.5cm}, clip]{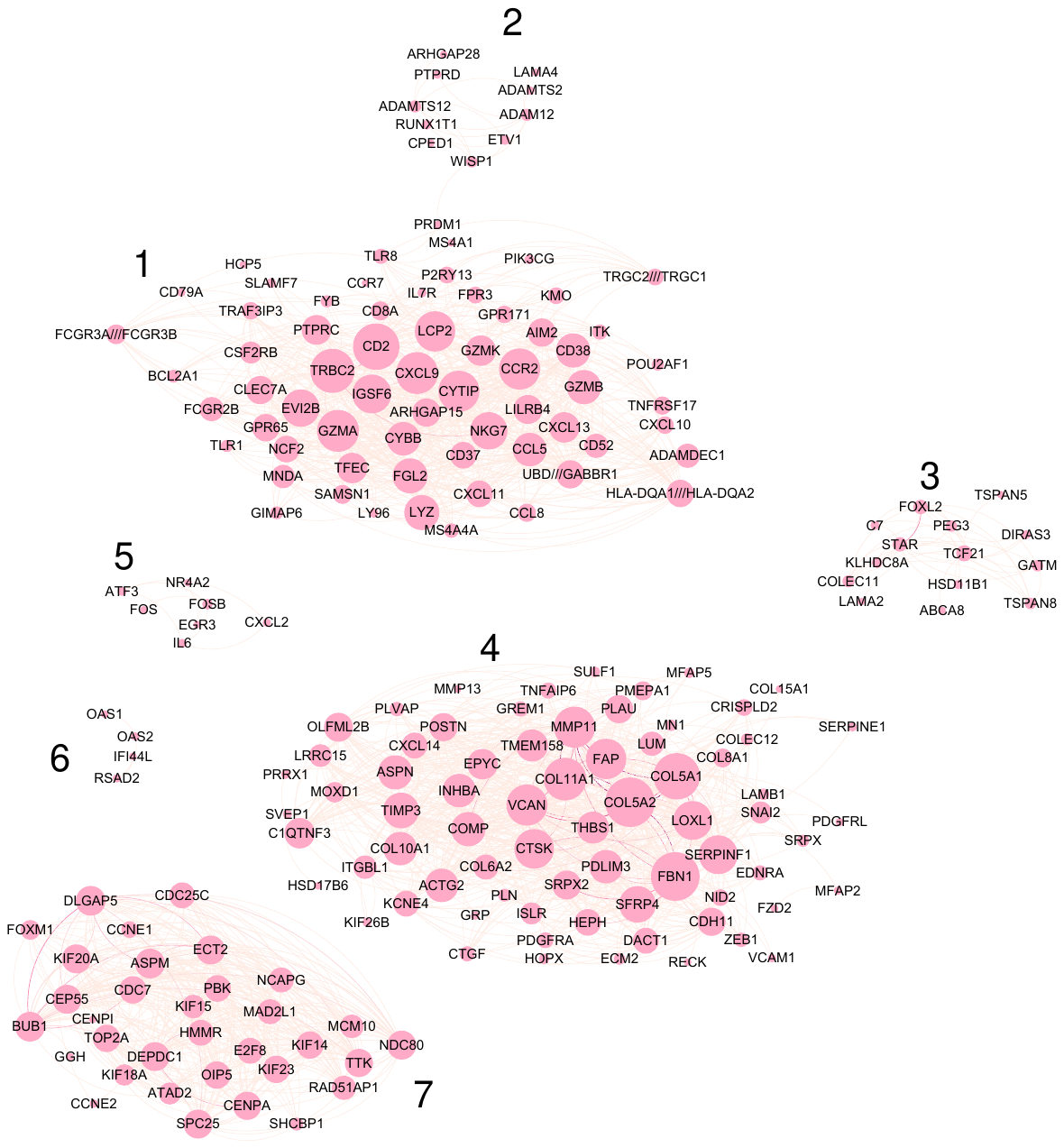}
    \label{fig:netPFA}
\end{subfigure}
\begin{subfigure}{.44\textwidth}
    \caption{SUFA}
    \centering
    \includegraphics[width=.95\linewidth, trim={0 0cm 0 0cm}, clip]{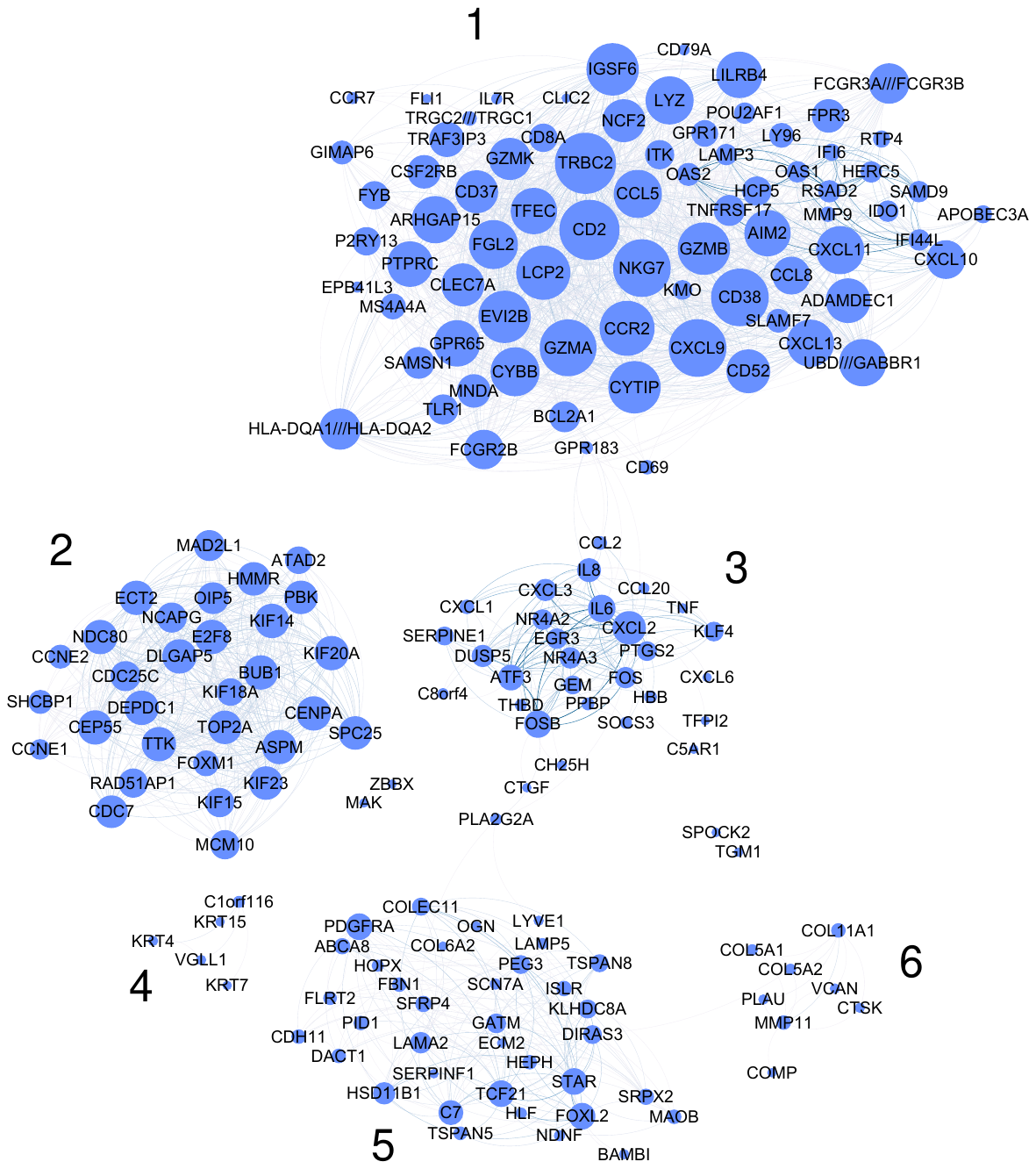}
    \label{fig:netSUFA}
\end{subfigure}\\
\begin{subfigure}{.47\textwidth}
    \centering
    \caption{BMSFA}
\includegraphics[width=.95\linewidth,trim={0 7cm 8cm 7cm}, clip]{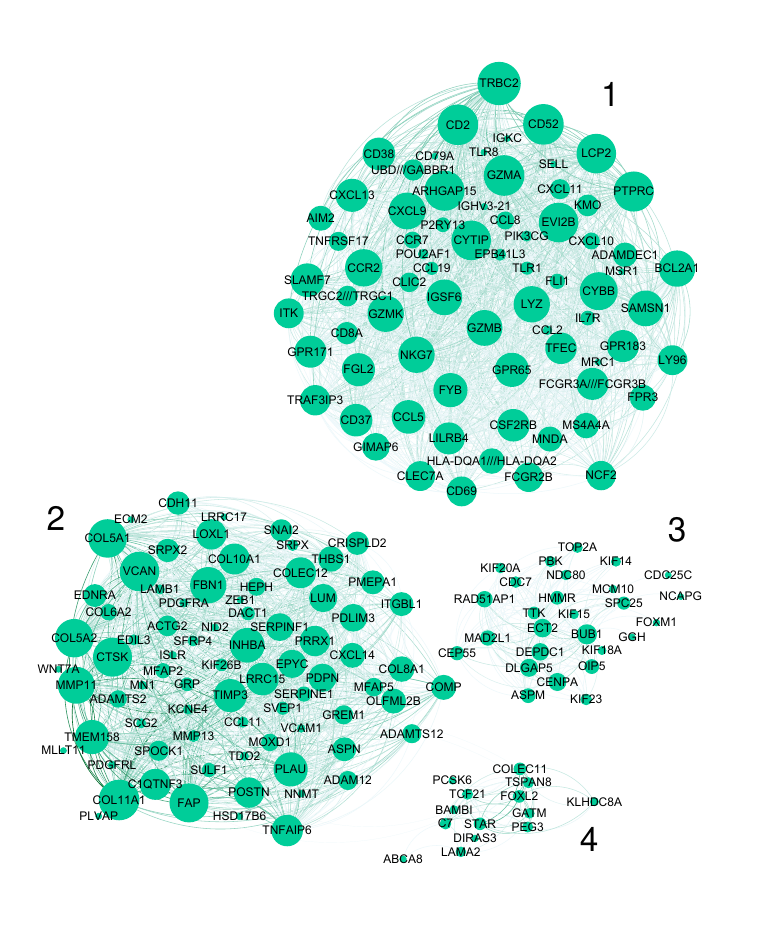}
    \label{fig:netBMSFA}
\end{subfigure}
\caption{\it Shared gene co-expression network based on $\widehat{\Sigma}_{\Phi}$ across the  ovarian cancer studies for different methods obtained using \texttt{Gephi} \cite{bastian2009gephi}. We include edges between two genes
if the corresponding element in the shared covariance matrix $\widehat{\Sigma}_{\Phi}
$ is greater than a threshold in absolute
value.  Numbers refer to clusters identified by the Yifan Hu\cite{hu2005efficient} algorithm.}
\label{fig:network}
\end{figure}

Figure \ref{fig:network} displays the shared gene co-expression network derived from the PFA, SUFA and BMSFA models.  Ind FA is excluded as it does not estimate common covariance. Stack FA and MOM-SS are also omitted from the figure because they produce overly dense networks: over 200 genes form a single large cluster due to uniformly high covariances (> 0.9). These results are included in the Supplementary Material. Tetris is excluded because the model did not complete within six days. Despite the long runtimes for PFA (4 days) and SUFA (over 10 hours), we include their results for comparative purposes.

In the co-expression network (Figure \ref{fig:network}), nodes represent genes, and edges correspond to the gene-gene covariances. Node size reflects the number of connections, and edge darkness reflects the magnitude of covariance.  Genes grouped in the same cluster may be co-regulated, share functional pathways, or reflect common transcriptional programs. These clusters highlight biological processes relevant to ovarian cancer progression and may help identify candidate prognostic markers \cite{sun2017gene}.

Cluster structure differs slightly across methods. BMSFA tends to group more genes into a single large cluster, while PFA and SUFA yield to several clusters. These differences can be related to  the distinct definitions of $\Sigma_\Phi$: in PFA, $\Sigma_\Phi$ is the marginal covariance of the first study (GSE13876\_eset); in SUFA, it is defined as $\Phi\Phi^\top + \Psi$; and in BMSFA, it corresponds to $\Phi\Phi^\top$. The number of extracted factors and estimation techniques further contribute to these structural differences. Nonetheless, several recurrent clusters across methods point to robust biological signal.

\textbf{Immune-related}: Cluster 1 in PFA, SUFA, and BMSFA includes key immune-regulatory genes such as \textit{CXCL9}, \textit{CXCL10}, \textit{CXCL11}, \textit{CXCL13}, and \textit{CCR2}. These chemokines and their receptors are critical in immune cell recruitment and activation,  mediating antitumor immunity \cite{breuer2013innatedb, tokunaga2018cxcl9, wang2019crosstalk}.  

\textbf{Extracellular matrix (ECM) organization related}: 
Cluster 4 (PFA), Cluster 6 (SUFA), and Cluster 2 (BMSFA) contain genes involved in extracellular matrix (ECM) organization and remodeling, including \textit{COL10A1}, \textit{COL11A1}, \textit{COL5A1}, \textit{POSTN}, \textit{VCAN}, \textit{TIMP3}, \textit{THBS1}, \textit{FAP}, and \textit{LOXL1}. These genes regulate ECM stiffness and integrity—factors known to influence tumor progression and metastasis \cite{giussani2018extracellular, lu2012extracellular}. 

\subsubsection{Model Prediction Accuracy}

As in the nutrition data analysis, model prediction accuracy is evaluated using mean squared error (MSE). Each dataset is randomly partitioned into a 70\% training set and a 30\% test set. Models are fit on the training data using the estimated number of factors reported in Table~\ref{tab:table_genomics}. Factor scores for the test set are then estimated using the fitted loadings and residual covariances, and used to reconstruct the observed test data. Tetris is excluded from this comparison, for high intensive computational time.

\begin{table}[htbp]
\centering
\caption{\it Estimated numbers of factors and MSE for each model on the gene expression data.}
\renewcommand{\arraystretch}{1.2}
\begin{tabular}{l c l l}
\toprule
\textbf{Model} & Estimated $K$ & Estimated $J_s$ & MSE \\
\midrule
Stack FA & 2 & - & 0.0208\\
Ind FA & - & (4,7,6,6) & 0.0143 \\
PFA & 20 & - & 0.0161 \\
MOM-SS & 20 & - & 0.0470\\
SUFA & 19 & (4, 4, 4, 4)  & 0.0131\\
BMSFA & 6 & (4, 4, 4, 4) & 0.0131\\
Tetris* & - & - & -\\
\bottomrule
\end{tabular} \\
\vspace{0.1cm}
Note: Tetris did not complete within 5 days. $K$: numbers of common factors;\\ $J_s$: numbers of study-specific factors.
\label{tab:table_genomics}
\end{table}

As shown in Table~\ref{tab:table_genomics}, all models demonstrate a good fit to the gene expression data. SUFA and BMSFA achieve the lowest MSE (0.0131). Although the MSE values are identical, BMSFA is preferable due to its more parsimonious factor structure—requiring fewer common factors than SUFA.

PFA also yields a competitive MSE but relies on a large number of factors ($K=20$), suggesting that the default cutoff parameter (0.001) may be too permissive for this dataset. Ind FA and BMSFA offer a strong balance between model complexity and prediction accuracy, demonstrating effective recovery of gene expression structure with relatively fewer factors.

\section{Discussion}\label{sec:5}

This tutorial presents a range of Bayesian integrative factor models for the analysis of multi-study, high-dimensional data. With the increasing volume and complexity of high-dimensional data generated across diverse studies and platforms, there is a critical need for advanced data integration techniques that can uncover shared and study-specific patterns in a principled manner. This demand is particularly pressing in fields such as genomics, nutrition, and epidemiology, where multi-study designs are now commonplace and require scalable, interpretable modeling strategies.

Through simulation and application to nutrition and genomics datasets, the comparative evaluation highlights trade-offs in flexibility, computational cost, and interpretability across models.


SUFA offers a favorable balance of accuracy and efficiency in small-to-moderate dimensions. BMSFA provides robust inference at the cost of additional runtime and post-processing. Tetris is appealing for exploratory analyses involving unknown latent structures but remains computationally intensive. MOM-SS enables rapid estimation, though with reduced accuracy in more complex settings. Stack FA and Ind FA serve as fast baselines but are limited in their capacity to capture both shared and study-specific structure. PFA is accurate but computationally demanding and may overfit in high-dimensional settings.

Each model exhibits distinct strengths. Stack FA assumes a shared factor structure across studies but ignores study-specific variation and tends to overestimate covariance, limiting its interpretability in network applications. Ind FA fits separate models per study, potentially capturing unique features, though at the expense of robustness due to limited sample sizes per study.

PFA introduces perturbation matrices to align studies to a reference, reducing rotational ambiguity and allowing straightforward post-processing. However, the indirect estimation of study-specific effects and the cost of computing full perturbation matrices pose challenges, particularly when $P$ is large.

MOM-SS jointly models covariates and latent factors using non-local spike-and-slab priors, yielding interpretable results with efficient EM-based inference. However, it can overestimate loadings and factor counts. SUFA explicitly separates common and study-specific factors and offers an efficient HMC sampler, though identifiability constraints may reduce precision in estimating study-specific effects. BMSFA relaxes these constraints, allowing full study-specific loadings, but requires a two-stage estimation procedure.

Tetris is the most flexible method, capturing fully shared, partially shared, and study-specific factors via a nonparametric prior. While offering automatic factor selection, it is highly computationally demanding and often infeasible for large-scale problems without substantial computing resources.

A central challenge remains the determination of the number of factors. Except for Tetris, all methods require a pre-specified or iteratively determined number. Stack FA, Ind FA, and BMSFA employ eigenvalue decomposition followed by model refitting, while MOM-SS and SUFA include procedures for adaptive selection, though they may lack precision. Balancing model complexity with interpretability is essential; overparameterization risks overfitting, while overly sparse models may obscure important structure.

Covariance matrices—such as $\Phi\Phi^\top$, $\Lambda_s\Lambda_s^\top$, and $\Psi_s$—are invariant to factor rotation and thus form a reliable basis for interpretation. These matrices support construction of interpretable networks, such as gene co-expression graphs or dietary pattern structures. In contrast, loading matrices require post-processing (e.g., varimax rotation or orthogonal Procrustes alignment) to resolve rotational non-identifiability. After rotation, sparsity in loadings enables interpretation of latent components, facilitating downstream tasks such as regression or survival analysis.

A practical workflow for Bayesian integrative factor analysis includes: (1) data preparation and preprocessing; (2) model selection and tuning; (3) estimation and post-processing for factor alignment and number determination; (4) interpretation via covariance networks or factor-based predictive models; and optionally (5) evaluation via predictive accuracy metrics such as MSE.
In summary, Bayesian integrative factor models provide a flexible framework for uncovering latent structure in multi-study settings. By balancing estimation accuracy, interpretability, and computational feasibility, these models enable principled integration and analysis across a wide range of domains. This tutorial aims to serve both as a practical guide for implementation and a foundation for future methodological innovation.


\section{Acknowledgments}

This Manuscript was prepared using HCHSSOL Research Materials obtained from the NHLBI Biologic Specimen and Data Repository Information Coordinating Center and does not necessarily reflect the opinions or views of the HCHSSOL or the NHLBI. 
The second real case study used in this work used data publicly available in the R package \texttt{curatedOvarianData}\cite{ganzfried2013curatedovariandata} available for download from Bioconductor at: \url{https://bioconductor.org/packages/release/data/experiment/html/curatedOvarianData.html}.
RDV was supported  by ``Programma per Giovani Ricercatori Rita Levi Montalcini'' granted by the Italian Ministry of Education, University, and Research.

\bibliographystyle{apalike}
\bibliography{ref}
\end{document}